\documentclass[prd,superscriptaddress,showpacs,floatfix,preprintnumbers, nofootinbib,hyperref]{revtex4}              
\usepackage{mathtools,ulem}
\usepackage{graphicx}
\usepackage{rotating}
\usepackage{epsfig}
\usepackage{amssymb}
\usepackage[lofdepth,lotdepth,caption=false]{subfig}

\usepackage{color}

\newcommand{\Ss}{{\mathcal S}}
\newcommand{\tr}{{\rm tr}}

\begin{document}

\title{Impact of sterile neutrinos on the early time flux from a galactic supernova}
\author{Arman Esmaili}
\affiliation{Instituto de F\'isica Gleb Wataghin - UNICAMP, 13083-859, Campinas, SP, Brazil}
\affiliation{Institute of Convergence Fundamental Studies, Seoul National University of Science and Technology, Gongreung-ro 232, Nowon-gu, Seoul 139-743, Korea}
\author{O. L. G. Peres}
\affiliation{Instituto de F\'isica Gleb Wataghin - UNICAMP, 13083-859, Campinas, SP, Brazil}
\affiliation{Abdus Salam International Centre for Theoretical Physics, ICTP, I-34010, Trieste, Italy}
\author{Pasquale Dario Serpico} 
\affiliation{LAPTh, Univ. de Savoie, CNRS, B.P.110, Annecy-le-Vieux F-74941, France}

\begin{abstract}
We study the impact of the existence of an eV-mass scale sterile neutrino---with parameters in the ballpark of what is required to fit the laboratory anomalies---on the early time profile of the electron neutrino and antineutrino fluxes associated to a core-collapse supernova (SN). In particular, we focus on the universal feature of neutronization burst expected in the first tens of ms of the signal: provided that a detector with sufficient sensitivity is available, it is well-known that in the 3 neutrino framework the detection of the neutronization burst in neutrino channel would signal inverted mass hierarchy. This conclusion is dramatically altered in the presence of a sterile neutrino: we study here both analytically and numerically the region in parameter space where this characteristic signal disappears, mimicking normal hierarchy expectations. Conversely, the detection of a peak consistent with expectations for inverted mass hierarchy would exclude the existence of a sterile state over a much wider parameter space than what required by laboratory anomalies fits, or even probed by detectors coming on-line in the near future. Additionally, we show the peculiar alteration in the energy-time double differential flux, with a delayed peak appearing for kinematical reasons, which might offer a remarkable signature in case of favorable parameters and for a high statistics detection of a Galactic SN. We also comment on additional potentially interesting effects in the electron antineutrino channel, if more than one angle in the active-sterile sector is non-vanishing. As an ancillary result that we derived in the technical resolution of the equations, in an appendix we report the Cayley-Hamilton formalism for the evolution of a four neutrino system in matter, generalizing existing results in the literature.
\end{abstract}

\date{\today}
\pacs{14.60.St,14.60.Pq,95.85.Ry\hfill LAPTH-005/14}
\maketitle

\section{Introduction}

A very exciting frontier of low-energy neutrino ($\nu$) astronomy is represented by the detection of neutrinos from core-collapse supernovae (SNe), whose first and till now only example was provided by SN1987A~\cite{Hirata:1987hu,Hirata:1988ad,Bionta:1987qt}. Existing large underground neutrino detectors (like SuperKamiokande or IceCube) as well as numerous planned ones are well suited to detect these rare galactic events (a few per century {\it in average}) with sufficiently high-statistics to allow for the extraction of detailed astrophysical information on the SN explosion mechanism~\cite{GilBotella:2004bv,Pagliaroli:2008ur,Pagliaroli:2009qy,Ianni:2009bd,Halzen:2009sm,Lund:2010kh}. Such measurements could also offer a handle on particle physics such as $\nu$ masses and mixings, too~\cite{Dighe:1999bi,Lunardini:2001pb,Lunardini:2003eh,Dighe:2003be,Abbasi:2011ss,Serpico:2011ir}.
  
One of the open questions in the neutrino sector is the existence of light sterile neutrino states, motivated by some experimental ``anomalies". Although a three (active) neutrino mixing scenario explains most of the data consistently, some short-baseline neutrino oscillation experiments suggest some deviations. These include the ${\overline \nu}_\mu \to {\overline\nu}_e$ oscillations in LSND~\cite{Aguilar:2001ty} and MiniBooNE~\cite{Aguilar-Arevalo:2013pmq} experiments, the ${\overline\nu}_e$ and $\nu_e$ disappearance dubbed Reactor Anomaly~\cite{Mention:2011rk} in reactor neutrino experiments and the Gallium Anomaly~\cite{Acero:2007su,Giunti:2010zu} in the calibration of solar neutrino experiments, respectively. Interestingly, a large fraction of these outlier data can be roughly accommodated if a light ($m\sim\mathcal O(1)\,$eV) sterile neutrino state is added to the picture (the so-called $3+1$ model)~\cite{Kopp:2013vaa,Giunti:2013aea,Giunti:2012tn}, although tensions between different sets of data persist. A plethora of experiments, relying on different strategies and methods, are in design or construction phase to check the existence of sterile neutrinos (see~\cite{Abazajian:2012ys} and references therein; see also~\cite{Esmaili:2013vza,Esmaili:2013cja,Esmaili:2012nz,Esmaili:2012vg}). 

Not surprisingly, the existence of such a sterile state would have an impact on SN neutrino conversions. Recently, for example, Ref.~\cite{Tamborra:2011is} showed how in such conditions the electron fraction $Y_e$ is reduced and the conditions for heavy-element formation in the supernova ejecta can be affected. It is also clear that the signal detectable at the Earth can be altered, see for instance~\cite{Choubey:2006aq,Choubey:2007ga} for an example associated to the turbulent shock wave.

In this article we discuss another interesting effect, which has passed almost unnoticed till now. In particular, the existence of sterile states with properties in the ball-park of what is required by the interpretation of ``anomalies" lead to interesting phenomenological consequences on the {\it neutronization burst}, that should be observable in a large detector of $\nu_e$. The neutronization burst is a prompt burst of $\nu_e$ associated to the passage of the newly formed shock to regions with densities low enough that neutrinos (initially trapped) begin to stream faster than the shock. Since the medium is basically made of free nucleons and is rich in $e^-$, the only rapid process is $e^-$ captures on $p$: this significantly suppresses the flux of flavors other than $\nu_e$, while the $\nu_e$ signal lasts. The existence and time profile of this burst are a generic feature of the early signal (first ${\cal O}$(20) ms post-bounce) of core-collapse SNe. Its properties are largely independent of the progenitor and still uncertain physical properties such as the dense matter equation of state (see for example~\cite{Kachelriess:2004ds} or Fig.~1 in~\cite{Serpico:2011ir}.) The detection of such a burst has already been discussed in the literature as a way to establish inverted mass hierarchy (IH) in the (active) neutrino sector~\cite{GilBotella:2003sz,Kachelriess:2004ds}, possibly  the most robust and spectacular one from SN $\nu$, provided that an instrument with enough sensitivity is available. In fact, for the presently measured ``large'' value of $\theta_{13}$, in the neutrino channel maximal $\nu_e$ conversion occurs for normal hierarchy (NH), while the survival probability is constant and given by $P_{ee}\simeq \sin^2 \theta_{12}\cos^2{\theta_{13}}\simeq 0.32$ for IH (see e.g.~\cite{GilBotella:2003sz} for details). Note that detectors such as a ${\cal O}(100)\,$kton Liquid Argon time projection chamber---of the same class proposed within the context of the LAGUNA collaboration for a future underground detector~\cite{Autiero:2007zj}---or a Megaton class water Cherenkov detector~\cite{Kachelriess:2004ds}, have already been shown to be capable of such a measurement of neutronization burst signal. 

This ``unambiguous'' picture is significantly altered in presence of a sterile neutrino with parameters fitting the laboratory anomalies. Notably, to first approximation the existence of sterile neutrino can make the $\nu_e$ burst signature disappear altogether! This fact has been mentioned in the past, see e.g.~\cite{Peres:2000ic,Murayama:2000hm}, but has never been studied in detail. Here we present a more precise analytical and numerical discussion of this signature: In particular, we identify the region in parameter space where the phenomenon takes place, comparing them to the preferred ones from sterile neutrino explanations of laboratory anomalies. Additionally, we present the peculiar alteration in the energy-time double differential flux, highlighting the presence of a delayed ``peak'' appearing for {\it kinematical} reasons, which may be a non-negligible feature for a sufficiently large mixing angle and mass in the sterile sector. The phenomenological importance of these features for diagnostics both in the active and sterile sector is discussed. We also study the consequences of assuming more than one non-vanishing angle in the active-sterile sector: this is particularly important for potential signatures in the $\bar{\nu}_e$ detection channel, which we briefly address.

This article is structured as follows: The formalism of $3+1$ scenario and discussions about the resonant active-sterile neutrino conversion is reported in Sec.~\ref{formalism}. In Sec.~\ref{FluxComp} we report the SN $\nu_e$ flux composition in $3+1$ scenario and in Sec.~\ref{sec:flux} we illustrate the effect of the existence of a sterile neutrino state with typical mixing parameters on representative time-energy double differential SN $\nu$ fluxes. Sec.~\ref{phenonuebar} is devoted to the SN $\overline{\nu}_e$ flux, its composition at Earth and phenomenological considerations. In Sec.~\ref{discussion} we shall discuss our results and finally conclude with some perspectives for forthcoming studies.

As a side remark of some technical importance, it is worth pointing out that we checked our analytical results (which assume factorization and adiabaticity) with a numerical code which implements the evolution of a $4\nu$ system with appropriate parameters in a (toy) SN matter potential, which is the generalization of the method described in~\cite{Ohlsson:1999xb} based on the Cayley-Hamilton formalism. Since we could not find the explicit result in the literature, we worked out the relevant formulae and report them in Appendix~\ref{4x4CH}. For completeness, in Appendix~\ref{FluxEarth} we also report the details of the derivation of our analytical results for the SN $\nu_e$ and $\overline{\nu}_e$ flux compositions outside the SN surface as a function of the input fluxes at the neutrinosphere.

\section{$3+1$ scenario: conversion probabilities and resonances\label{formalism}}

In this section we discuss the impact of sterile neutrinos on the SN neutrino flux. Subsection~\ref{prem} summarizes the results of numerical simulation of SN explosion and its expected neutrino flux. In subsection~\ref{formalismVac} we briefly discuss the mixing in neutrino sector in the presence of one sterile neutrino (the $3+1$ scenario) and current best-fit values of active-sterile mixing parameters. In subsection~\ref{formalismMatt} we study in detail the oscillation of neutrinos in $3+1$ scenario in the medium of SNe.     

\subsection{Preparatory Materials\label{prem}}

Before discussing the effect of sterile neutrino on SN flux, in this subsection we summarize some basic information on the neutrino and anti-neutrino SN fluxes. 
Numerical simulations of core-collapse SNe provide the un-oscillated doubly differential neutrino distribution in energy and time,
\begin{equation}
F^0_\nu (E_\nu , t)\equiv\frac{d^2 N_\nu}{d t \,d E_\nu}~,\label{eq:nudistrib}
\end{equation}
where $\nu=\left\{ \nu_e, \overline\nu_e, \nu_x\right\}$ in the standard notation~\cite{Dighe:1999bi}. This is related to the instantaneous (time-dependent) luminosity via
\begin{equation}\label{eq:lum}
L_\nu=\int_{0}^{\infty} dE_\nu E_\nu F^0_\nu  \,\ .
\end{equation}

\begin{figure}[t!]
\centering
 \includegraphics[trim= 0mm 0mm 0mm 
0mm,clip,width=0.7\textwidth]{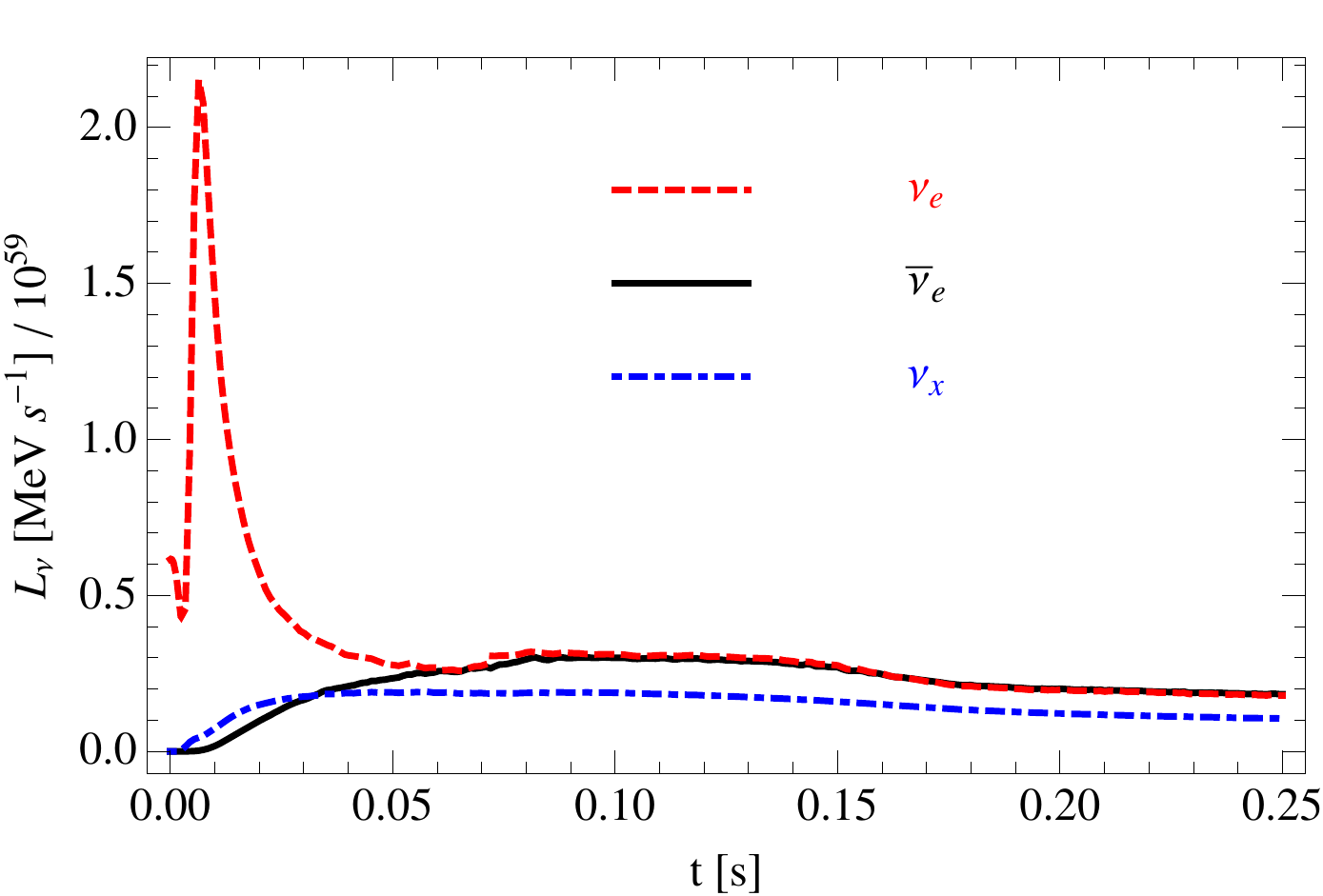}
\caption{\label{fig:lum}The luminosity of $\nu_e$, $\bar{\nu}_e$ and $\nu_x$ at production region, given in Eq.~(\ref{eq:lum}), from Garching simulation~\cite{Garchingmodels} of a $20M_\odot$ progenitor SN~\cite{Woosley:2002zz}.}
\end{figure}

We factorize simulation outputs as follows:
\begin{equation}
F^0_\nu (E_\nu , t)= \frac{dN_\nu}{dt}\varphi(E_\nu)~,
\end{equation}
for each flavor ($\nu=\nu_e, \overline\nu_e, \nu_x$), where
\begin{equation}
\frac{dN_\nu}{dt} = \frac{L_\nu}{\langle E_\nu\rangle}~,
\end{equation}
represents the neutrino emission rate (number of $\nu$'s per unit of time) with mean neutrino energy $\langle E_\nu\rangle$. The function $\varphi(E_\nu)$ is the normalized ($\int \varphi(E_\nu)dE_\nu = 1$) energy spectrum parametrized as in~\cite{Keil:2002in}
\begin{equation}
\varphi(E_\nu)= \frac{1}{\langle E_\nu\rangle}\frac{(1+\alpha)^{1+\alpha}}{\Gamma(1+\alpha)}\left(\frac{E_\nu}{\langle E_\nu\rangle}\right)^\alpha\exp\left[-(1+\alpha)\frac{E_\nu}{\langle
E_\nu\rangle}\right]\, ,
 \label{eq:varphi}
\end{equation}
where the energy-shape parameter $\alpha$ is defined as~\cite{Keil:2002in,Raffelt:2001}
\begin{equation}
\alpha=\frac{2\langle E_\nu \rangle^2-\langle E_\nu^2\rangle}{\langle E_\nu^2\rangle-
\langle E_\nu\rangle^2} \,, \label{alphadef}
\end{equation}
i.e. it is a dimensionless parameter containing information on the second moment of the distribution, $\langle E_\nu^2\rangle$. In general, $L_\nu$, $\langle E_\nu\rangle$ and $\alpha$ are all functions of time, and are extracted directly from the simulations. For definiteness, in this paper we use as benchmark the spherically symmetric Garching simulation~\cite{Garchingmodels} of a $20M_\odot$ progenitor SN from~\cite{Woosley:2002zz}, focusing our attention on post-bounce times $t<250$~ms. Figure~\ref{fig:lum} shows the time evolution of the luminosity of un-oscillated neutrino and anti-neutrino fluxes at production region given in Eq.~(\ref{eq:lum}). The peak in $\nu_e$ flux (the red dashed curve in Figure~\ref{fig:lum}) is the neutronization burst. Note that several studies have established that the properties of the neutronization burst are largely independent of the progenitor and still uncertain physical properties such as the dense matter equation of state, and its normalization is so robust that it has even been proposed as a ``standard candle'' for a SN distance determination~\cite{Kachelriess:2004ds}.

\subsection{$3+1$ Scenario Formalism in Vacuum and Numerical Approach in Matter\label{formalismVac}}

In a four-neutrino mixing scheme (the so-called $3+1$ scenario), the flavor neutrino basis is composed of the three active neutrinos $\nu_e, \nu_{\mu}, \nu_{\tau}$ and a sterile neutrino $\nu_s$. The flavor eigenstates $\nu_\alpha$ are related to the mass eigenstates $\nu_i$ ($i=1,\ldots 4$, ordered by growing mass) via a unitary matrix ${\mathcal U}$ through
\begin{equation} \nu_{\alpha} = {\mathcal U}_{\alpha i}^* \, \nu_i~, \qquad {\rm where} \qquad {\mathcal U}\,{\mathcal U}^{\dagger}= {\mathcal
U}^{\dagger}\,{\mathcal U}= {\sf I}~. \label{nualphanui}
\end{equation}
Different parameterizations are possible for the matrix $\, {\mathcal U}$; for example, it can parameterized as a product of Euler rotation matrices $R_{ij}$ acting in the $(i,j)$ mass eigenstates subspace, each specified by a mixing angle $\theta_{ij}$. Thus, one can write
\begin{equation} {\mathcal U}= R_{34}R_{24}R_{23}R_{14}R_{13}R_{12}~,
\end{equation}
where the flavor eigenstates are ordered in such a way that if all the mixing angles vanish we have the correspondence $(\nu_e,\nu_\mu,\nu_\tau,\nu_s)=(\nu_1,\nu_2,\nu_3, \nu_4)$, for the NH case among active neutrino. In the limit where the three mixing angles $\theta_{i4}$ ($i=1,2,3$) vanish, the above matrix reduces to
\begin{equation}
\lim_{\theta_{i4}\to 0}{\mathcal U} = 
\left(\begin{array}{cc} 
U(\theta_{12}, \theta_{13}, \theta_{23}) &0 \\
0 & 1 \\
\end{array}\right)~,
\label{eq:matrix}
\end{equation}
where $U$ is the conventional $3 \times 3$ unitary mixing matrix (PMNS matrix) among the active neutrinos defined in terms of three rotation angles $(\theta_{12},\theta_{23},\theta_{13})$. In the following we shall assume that ${\mathcal U}$ is real, and we shall fix the mixing angles entering $R_{23}$, $R_{13}$, and $R_{12}$ to the best-fit values from a global analysis of oscillation data~\cite{GonzalezGarcia:2012sz} (see also~\cite{Fogli:2012ua,Tortola:2012te}) 
\begin{equation}
\sin^2 \theta_{12} = 0.3 \quad , \quad \sin^2 \theta_{23} = 0.5 \quad , \quad  \sin^2  \theta_{13} = 0.023~.
\end{equation}
For future reference, in the limit of a vanishing $\theta_{14}$ one has $|U_{e2}|^2=\sin^2\theta_{12}\cos^2\theta_{13}$ and $|U_{e3}|^2= \sin^2  \theta_{13}$ and these two matrix elements are independent of $\theta_{24}$ and $\theta_{34}$. Notice that reactor and Gallium anomalies favor a nonzero $\theta_{14}$; while the LSND/MiniBooNE anomaly requires both $\theta_{14}\neq0$ and $\theta_{24}\neq0$. The angle $\theta_{34}$ is the least constrained active-sterile mixing angle and can be put to zero in the interpretation of anomalies. It is shown in~\cite{Esmaili:2013cja} that IceCube can constrain this angle to a level comparable to the other angles. The global analysis of the above-mentioned ``anomalies" in $3+1$ scenario leads to the following best-fit values for $\theta_{14}$ and $\theta_{24}$ mixing angles (taken from~\cite{Kopp:2013vaa,Giunti:2013aea}) 
\begin{equation}\label{eq:active-sterile}
\sin^2 \theta_{14} = 0.023 \quad , \quad \sin^2 \theta_{24} = 0.029~.
\end{equation}
In the calculations of the rest of this paper, when a nonzero value for active-sterile mixing angles are considered, we use the values of Eq.~(\ref{eq:active-sterile}) as the benchmark.

The expressions of transition and survival probabilities $P(\nu_\alpha\to \nu_\beta)$ are cumbersome but straightforward to obtain analytically in the case of pure vacuum oscillations. It is well-known (see e.g.~\cite{Dighe:1999bi}) that for neutrinos propagating out of SN core the vacuum approximation is far from being sufficient, since a relevant role is played by the matter refractive potential in the stellar envelope, which induces the celebrated Mikheyev-Smirnov-Wolfenstein~(MSW) effect~\cite{Matt}. Note that it has been recently realized that in the deepest SN regions the neutrino density is so high that the neutrino-neutrino interactions~\cite{Pantaleone:1992eq,Qian:1994wh} may dominate the flavor evolution in a highly non-trivial way (for a review see~\cite{Duan:2010bg}). Fascinating (and hierarchy dependent) collective phenomena causing coherent conversions with peculiar energy dependences of the type $\nu_x \bar{\nu}_x \leftrightarrow \nu_e\bar{\nu}_e$ have been uncovered, but we caution the reader that they are only partially understood and have been modeled under a number of simplifications, so that results concerning those effects have to be taken as preliminary. Nonetheless, for the early time signal of interest here such effects are either almost absent in principle, since during the neutronization burst there are small fluxes of antineutrinos or, concerning the slightly longer timescales of few hundreds ms of the accretion phase, they are typically found to be suppressed by multi-angle ``matter'' effects~\cite{EstebanPretel:2008ni,Chakraborty:2011nf,Chakraborty:2011gd}, at least for massive enough progenitors. Hence we neglect them in the following.

The evolution of the neutrino state $\nu(r)= (\nu_e(r), \nu_\mu(r),\nu_\tau(r),\nu_s(r))^T$ is written in terms of the fluxes at the neutrinosphere ($r_0$) as
\begin{equation}\label{eq:smatrix} 
\nu(r) = \Ss(r) \nu(r_0)~,
\end{equation} 
where the evolution operator $\Ss(r)$ depends on the distance $r$ traversed in the medium and the medium properties. In terms of $\Ss$, the probability of an initial neutrino of flavor $\nu_\alpha$ to be in the flavor eigenstate $\nu_\beta$ at $r$ is then given by $P(\nu_{\alpha}\to \nu_{\beta};r)\equiv P_{\alpha\beta}(r)  = |\Ss_{\beta\alpha}(r)|^2$. Once the flavor composition at the exit of the SN is known, the flux in the mass basis can be simply obtained by inverting Eq.~(\ref{nualphanui}), which will be the same as on the Earth. The details of the calculation of the evolution operator, which is used for the numerical resolution of the system, are given in the Appendix~\ref{4x4CH}. They involve a generalization of the method described in~\cite{Ohlsson:1999xb}, with some technicalities worth reporting separately. 

\subsection{Active-Sterile Conversion Probabilities: Resonances\label{formalismMatt}}

Although all results we are interested in can be obtained numerically, the basic physics of the flavor conversion leading to active-sterile conversion can be grasped analytically, as we describe in the following. For our purposes, it suffices to approximate a typical matter density profile as (for post-bounce times $<1$~s, see e.g.~\cite{Schirato:2002tg,Fogli:2003dw})
\begin{equation}\label{eq:rho}
\rho(x)\approx 10^{14} \left(\frac{x}{\rm km}\right)^{-2.4} \,{\rm g/cm}^3\:\:\:(x\gtrsim 10\,{\rm km})~.
\end{equation}
In the $\{\nu_e,\nu_x,\nu_s\}$ system (where $x=\mu$ and $\tau$), the matter potential writes
\begin{equation}
V=\sqrt{2}G_{\rm F}\left(N_e-N_n/2,-N_n/2,0\right)= V_{CC} \left(1-\frac{N_n}{2\,N_e},-\frac{N_n}{2\,N_e},0\right)\,,
\end{equation}
where $N_e$ and $N_n$ are the electron and neutron number densities, respectively; and $V_{CC}\equiv \sqrt{2}G_{\rm F}N_e$. In terms of the electron fraction $Y_e$,
\begin{equation}
\frac{N_n}{N_e}=\frac{1}{Y_e}-1\,.
\end{equation}
A typical value is $Y_e\approx 0.5$, hence
\begin{equation}
V=V_{CC} \left(\frac{3}{2}-\frac{1}{2\,Y_e},\frac{1}{2}-\frac{1}{2Y_e},0\right)\,\approx V_{CC}(0.5,-0.5,0)~,
\end{equation}
with the pre-factor $V_{CC}$ writing in convenient units in terms of $\rho(x)$ as
\begin{equation}
V_{CC}= 7.6\times 10^{-8}\;Y_e\,\frac{\rho(x)}{{\rm g/cm}^3}\,\frac{\rm eV^2}{\rm MeV}\,.
\end{equation}

\begin{figure}[!ht]
\centering
\subfloat[$P(\nu_e\to\nu_s)$, both NH and IH]{
 \includegraphics[trim= 0mm 0mm 0mm 
80mm,clip,width=0.45\textwidth]{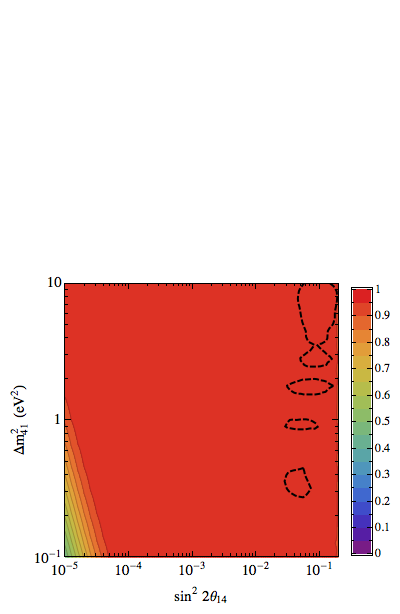}
  \label{fig:etos-high}
}
\subfloat[$p_{\rm jump}=\exp (-\pi\gamma /2)$]{
 \includegraphics[trim= 0mm 0mm 0mm 
80mm,clip,width=0.45\textwidth]{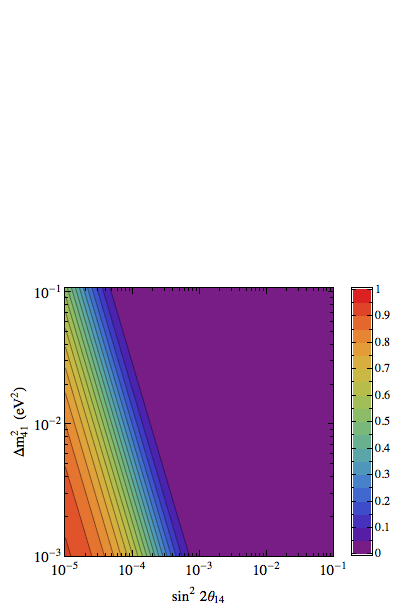}
  \label{fig:pjump}
}
\quad
\subfloat[$P(\nu_e\to\nu_s)$, NH]{
 \includegraphics[trim= 0mm 0mm 0mm 
80mm,clip,width=0.45\textwidth]{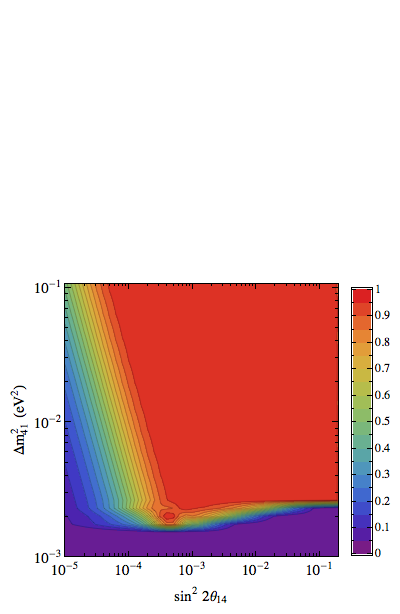}
  \label{fig:etos-prob}
}
\subfloat[$P(\nu_e\to\nu_s)$, IH]{
 \includegraphics[trim= 0mm 0mm 0mm 
80mm,clip,width=0.45\textwidth]{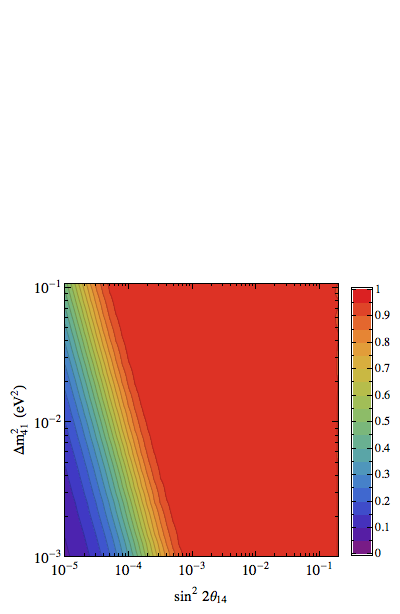}
  \label{fig:etos-probIH}
}
\caption{\label{fig:fig1} Panel (a) shows the conversion probability $P(\nu_e\to\nu_s)$ for large values of $\Delta m_{41}^2$ for both NH and IH. Panel (b) shows the jumping probability $p_{\rm jump}$. Panel (c) shows $P(\nu_e\to\nu_s)$ for small values of $\Delta m_{41}^2$ assuming Normal Hierarchy of active neutrinos, while panel (d) is the same for Inverted Hierarchy. The black dashed curves in panel (a) shows the allowed region from the global analysis of all short baseline disappearance data at 95\% C.L.~\cite{Kopp:2013vaa}. For all the panels we assume $E_\nu=10$~MeV and $\theta_{24}=\theta_{34}=0$.}
\end{figure}

Very deep in the SN mantle the electron fraction $Y_e<1/3$ and the potential is negative for $\nu_e$, but the corresponding transition probabilities are extremely non-adiabatic (see e.g.~\cite{Choubey:2007ga}) and we shall ignore them in the following. Under this assumption and the further 
 $2\nu$ approximation for the resonance description,  the resonance condition for the $\nu_e-\nu_s$ conversion can be written as:
\begin{equation}\label{eq:es-res} \frac{\Delta m_{41}^2
\cos2\theta_{14}}{2E_\nu} = \frac{\sqrt{2}G_F\rho}{2m_N}~,
\end{equation} 
where $m_N$ is the total nucleon mass; that is $m_N\approx m_n+m_p$. Notice that we are assuming $Y_e=0.5$, hence $N_e=N_p=N_n$. Remembering that $\Delta m_{21}^2\simeq 8\times10^{-5}\,$eV$^2$ and $|\Delta m_{31}^2|\simeq 2\times10^{-3}\,$eV$^2$, the short-baseline experiment hints for sterile neutrinos require $|\Delta m_{41}^2|\gg |\Delta m_{ji}^2|,\,i<j\leq 3$ (see also the dashed contours in Fig.~\ref{fig:etos-high}). So, $\Delta m_{41}^2<0$ would imply that all four neutrino states have an absolute mass of the same order of $\sqrt{|\Delta m_{41}^2|}$. This would lead to severe conflict with cosmological bounds~\cite{Ade:2013zuv}, and---in a part of the parameter space---also with the direct bound from tritium beta decay~\cite{Kraus:2004zw}. Hence, in this paper we always assume $\Delta m_{41}^2 > 0$. However, we will consider a broader parameter space than the one hinted to by laboratory anomalies. The effects discussed in this article are in fact relevant in a wider range of mass and mixing angle parameters, which we want to characterize. Note also that, while fitting laboratory anomalies is accompanied by some tension with cosmological data, lighter and/or more weakly coupled sterile neutrinos could also {\it improve} the cosmological fits, as discussed for instance in~\cite{Hamann:2013iba}. Some of the phenomena described in the following provide perhaps the unique viable check of this broader parameter space.

From Eq.~(\ref{eq:es-res}), for a fixed value of $E_\nu$ the part of $(\Delta m_{41}^2,\sin^22\theta_{14})$ parameter space for which the resonance occurs can be determined. Assuming that the radius of neutrinosphere is $\sim30$~km, it is straightforward to show that the resonance occurs for
\begin{equation}\label{resupp} \Delta m_{41}^2 \cos2\theta_{14}\lesssim 10^4~{\rm eV}^2 \left( \frac{E_\nu}{10~{\rm MeV}} \right)~.
\end{equation} 
However, the resonance is not adiabatic for all the values of $\Delta m_{41}^2$. For a density profile $\rho(r)=A r^{-\eta}$ the adiabaticity parameter $\gamma$ is given by
\begin{equation} \gamma = \frac{\Delta m_{41}^2}{2E_\nu}
\frac{\sin^22\theta_{14}}{\cos2\theta_{14}}
\frac{1}{\left|\frac{1}{N}\frac{dN}{dr}\right|_{\rm res}} =
\frac{1}{2\eta} \left( \frac{\Delta m_{41}^2}{E_\nu}
\right)^{1-1/\eta}
\frac{\sin^22\theta_{14}}{\left(\cos2\theta_{14}\right)^{1+1/\eta}}
\left( \frac{\sqrt{2}AG_F}{m_N} \right)^{1/\eta}~.
\end{equation} 
The jumping probability (level-crossing) at resonance region is\footnote{The probability $p_{\rm jump} = \exp(-\pi\gamma/2)$ is for densities with linear position dependence. For $r^{-\eta}$ dependence the jumping probability is given by $p_{\rm jump} = \exp(-F\pi\gamma/2)$, where $F =2 \sum_{m=0}^{\infty} C(-1/\eta-1,2m) C(1/2,m+1) (\tan2\theta_{14})^{2m}$ with $C$ denoting the binomial coefficient~\cite{Kuo:1989qe,Kuo:1988pn}. However, for $\eta=2.4$ and small mixing angle $\theta_{14}$ we have $F\simeq1$.} $p_{\rm jump} \approx \exp(-\pi\gamma/2)$, and is depicted in Fig.~\ref{fig:pjump}. Assuming the density profile of Eq.~(\ref{eq:rho}) with $\eta=2.4$, the adiabaticity parameter takes the values
\begin{equation}\label{eq:gamma} \gamma \simeq 10^2 \left(
\frac{\Delta m_{41}^2}{10^{-2}~{\rm eV}^2}
\right)^{\frac{\eta-1}{\eta}} \left( \frac{E_\nu}{10~{\rm MeV}}
\right)^{\frac{1-\eta}{\eta}} \left(
\frac{\sin^22\theta_{14}}{10^{-2}} \right)~~~~,~~~~ {\rm
for}\quad\eta=2.4 \quad {\rm and}\quad \theta_{14}\ll1\,.
\end{equation} 
From the adiabaticity condition ($\gamma \gtrsim 1$) a lower bound on $\Delta m_{41}^2$ can be derived, which of course depends on $\sin^22\theta_{14}$ and $E_\nu$. However, it should be noticed that the $\gamma$-factor in Eq.~(\ref{eq:gamma}) is obtained by assuming factorization of dynamics near level crossing zones corresponding to $\Delta m_{41}^2$, $\Delta m_{31}^2$ and $\Delta m_{21}^2$. Obviously this factorization assumption breaks down for very small values of $\Delta m_{41}^2$. Two cases can be identified: 

(i) For the normal hierarchy (NH) ordering between the active states, the $\nu_e$ produced in the supernova (which is $\nu_{4m}$, $m$ denoting the instantaneous mass eigenstate in matter) propagate adiabatically out of the supernova if $\Delta m_{41}^2$ is in the following range
\begin{equation}
\label{eq:NHrange} 
\max \left[ \frac{\Delta m_{31}^2}{{\rm eV}^2}, 10^{\frac{2-6\eta}{\eta-1}} \left(
\frac{E_\nu}{10~{\rm MeV}} \right) \left( \sin^22\theta_{14}
\right)^{\frac{\eta}{1-\eta}} \right] \lesssim \frac{\Delta
m_{41}^2}{{\rm eV}^2} \lesssim 10^4 \left( \frac{E_\nu}{10~{\rm MeV}}
\right) \left( \frac{1}{\cos2\theta_{14}} \right)~.
\end{equation} 
Thus, provided that Eq.~(\ref{eq:NHrange}) is satisfied, a complete conversion of $\nu_e\to\nu_4$ occurs which leads to the probabilities $P(\nu_e\to\nu_s)=|U_{s4}|^2$ and $P(\nu_e\to\nu_e)=|U_{e4}|^2$. The lower limit in Eq.~(\ref{eq:NHrange}) comes from the fact that when $\Delta m_{41}^2\simeq\Delta m_{31}^2$ the corresponding two level crossing zones merge and factorization is not possible anymore. For $\Delta m_{41}^2\lesssim \Delta m_{31}^2$ (and still adiabatic propagation) we obtain a complete conversion of $\nu_e\to\nu_3$ and so $P(\nu_e\to\nu_s)=|U_{s3}|^2$ and $P(\nu_e\to\nu_e)=|U_{e3}|^2$. So in this case, although the $P(\nu_e\to\nu_s)$ is small, the $\nu_e$ flux converts almost completely to $\nu_\mu$ and $\nu_\tau$ and the neutronization burst disappears in $\nu_e$ channel. 

(ii) For the inverted hierarchy (IH) ordering of active neutrinos the resonance due to the $\Delta m_{31}^2$ splitting is in the anti-neutrino channel; a complete conversion of $\nu_e\to\nu_s$ occurs for $\Delta m_{41}^2$ in the following range
\begin{equation}\label{eq:IHrange} 
\max \left[ \frac{\Delta m_{21}^2}{{\rm eV}^2}, 10^{\frac{2-6\eta}{\eta-1}} \left(
\frac{E_\nu}{10~{\rm MeV}} \right) \left( \sin^22\theta_{14}
\right)^{\frac{\eta}{1-\eta}} \right] \lesssim \frac{\Delta
m_{41}^2}{{\rm eV}^2} \lesssim 10^4 \left( \frac{E_\nu}{10~{\rm MeV}}
\right) \left( \frac{1}{\cos2\theta_{14}} \right)~.
\end{equation} 
In the range of Eq.~(\ref{eq:IHrange}), we obtain again $P(\nu_e\to\nu_s)=|U_{s4}|^2$ and $P(\nu_e\to\nu_e)=|U_{e4}|^2$. It is only for much smaller splitting, $\Delta m_{41}^2\lesssim \Delta m_{21}^2$, that the $\nu_e$ state in the deep part of supernova almost completely converts to $\nu_2$ during the propagation out of the supernova and $P(\nu_e\to\nu_s)=|U_{s2}|^2$ and $P(\nu_e\to\nu_e)=|U_{e2}|^2$.

The above discussion can be straightforwardly generalized to the case where $\theta_{24}\neq0$ and/or $\theta_{34}\neq0$. Nonzero values of $\theta_{24}$ and $\theta_{34}$ lead to resonant conversion of $\bar{\nu}_\mu\to\bar{\nu}_4$ and $\bar{\nu}_\tau\to\bar{\nu}_4$ respectively, which do not affect the neutronization burst flux. However, non-vanishing $\theta_{24}$ and $\theta_{34}$ change the values of $|U_{s2}|^2$ and $|U_{s3}|^2$.

To illustrate the cases (i) and (ii) we show in Figures~\ref{fig:etos-high}, \ref{fig:etos-prob} and \ref{fig:etos-probIH} the probability of $\nu_e\to\nu_s$ conversion in the plane $(\sin^22\theta_{14},\Delta m_{41}^2)$ for the fixed neutrino energy $E_\nu=10$~MeV (assuming $\theta_{24}=\theta_{34}=0$). For better visibility, we split the broad range of $\Delta m_{41}^2$ into small values in Figures~\ref{fig:etos-prob} and \ref{fig:etos-probIH}, respectively for NH and IH; and larger values in Figure~\ref{fig:etos-high} for both NH and IH. Also, Figure~\ref{fig:pjump} shows the $p_{\rm jump}$ for the same energy $E_\nu=10$~MeV and small values of $\Delta m_{41}^2$. All the panels of Figure~\ref{fig:fig1} are calculated numerically. As can be seen, for NH in Figure~\ref{fig:etos-prob}, the $\nu_e-\nu_s$ resonance is adiabatic for $\Delta m_{41}^2\gtrsim \Delta m_{31}^2$ and $\sin^22\theta_{14}\gtrsim 5\times 10^{-3}$. For $\Delta m_{41}^2\lesssim \Delta m_{31}^2$ in Figure~\ref{fig:etos-prob} the $\nu_e$ converts to $\nu_3$ and so $P(\nu_e\to\nu_s)=|U_{s3}|^2=\sin^2 \theta_{13} \sin^2\theta_{14}$ for $\theta_{24}=\theta_{34}=0$; which is quite small. However, for non-vanishing $\theta_{24}$ and $\theta_{34}$, the matrix element $U_{s3}$ can be as large as $(\cos\theta_{34}\sin\theta_{24}+\sin\theta_{34})/\sqrt{2}$; which by considering the current upper limits can lead to $\nu_e\to\nu_s$ oscillation probability as large as $\sim 0.2$. For IH in Figure~\ref{fig:etos-probIH}, the $P(\nu_e\to\nu_s)$ oscillogram mimics the same pattern as jumping probability in Figure~\ref{fig:pjump} for $\Delta m_{41}^2$ down to $\Delta m_{21}^2$ (not visibile in the figure). In the $\Delta m_{41}^2\lesssim \Delta m_{21}^2$ region, $\nu_e$ converts to $\nu_2$ and $P(\nu_e\to\nu_s)=|U_{s2}|^2\lesssim0.05$ from current upper limits.

In particular, we note that for values of $\Delta m_{41}^2$ motivated by the ``reactor anomaly'', as shown in Figure~\ref{fig:etos-high} by black dashed curves, one has $P(\nu_e\to\nu_s)=|U_{s4}|^2$ for $\sin^22\theta_{14}\gtrsim 10^{-5}$ for both NH and IH. Also, for $\Delta m_{41}^2 \sim 10^{-2}~{\rm eV}^2$ and $\sin^22\theta_{14}\sim0.06$ (suggested e.g. in~\cite{Esmaili:2013yea} for the interpretation of medium baseline reactor experiments) $\nu_e\to\nu_s$ conversion takes place adiabatically. These are in agreement with Eqs.~(\ref{eq:NHrange}) and (\ref{eq:IHrange}).

\section{SN $\nu_e$ flux in $3+1$ scenario\label{pheno}}

\subsection{$\nu_e$ Flux Composition at the Earth}\label{FluxComp}

Of course, the flux evolution is altered further (in a way that depends on the pattern of mass hierarchy in the active sector) when neutrinos cross the ``lower densities'' resonances. Eventually, the flux composition at the exit of the SN will be given by a linear combination of the initial fluxes as
\begin{eqnarray}
\label{eq:nueflux} 
F_{\nu_e} = c_{ee} F^0_{\nu_e} +
c_{xe} F^0_{\nu_x} +  c_{se} F^0_{\nu_s}\,.
\end{eqnarray} 
Here for completeness we consider a possible non-vanishing initial flux of sterile neutrinos, although we put $F_{\nu_s}^{0}=0$ in the following numerical evaluations. The expressions for the coefficients $c_{ij}$ in the standard 3$\nu$ scenario are well-known in the literature~\cite{Dighe:1999bi} and are reported in the left part of Table~\ref{tabI}. In the $3+1$ framework they are obviously modified (see Appendix~\ref{FluxEarth} for the explicit derivation). Their analytical expressions in the limiting case where all resonances are factorized and adiabatic are given in the last columns of Table~\ref{tabI}. Also numerical values for a benchmark value of $\theta_{14}$ are shown in Table~\ref{tabI}. All these results were checked numerically and were found to agree within the significant digits reported in the table and often better; typical discrepancies only arise at the $\sim10^{-3}$ level or below, where we are limited anyway by the numerical errors. Also, it is worth noticing that the $\nu_e$ flux in Eq.~(\ref{eq:nueflux}) only depends on $\theta_{14}$ active-sterile mixing angle, as long as $\Delta m_{41}^2$ falls in the range of Eqs.~(\ref{eq:NHrange}) and (\ref{eq:IHrange}), and so is independent of $\theta_{24}$ and $\theta_{34}$.

\begin{table}[!htb]
\caption{Coefficients in Eq.~(\ref{eq:nueflux}) in the 3$\nu$ and $3+1$ frameworks for both NH and IH. The analytical expressions are valid in the whole parameter space of $3+1$ scenario, including $\theta_{24}\neq0$ and/or $\theta_{34}\neq0$. The reported numerical values are for mixing angle values: $\theta_{14}=8.7^\circ$ (best-fit value from~\cite{Kopp:2013vaa,Giunti:2013aea}), $\theta_{24}=\theta_{34}=0$. The oscillation parameters in the (sub)matrix $U$ of Eq.~(\ref{eq:matrix}) are fixed to the best-fit values from global analysis of oscillation data~\cite{GonzalezGarcia:2012sz}: $\theta_{12}=33^\circ$, $\theta_{23}=45^\circ$ and $\theta_{13}=8.7^\circ$.}\begin{center}
\begin{tabular}{|c|c|c|c|c|}
\cline{2-3}\cline{4-5}
\cline{2-3}\cline{4-5}
\multicolumn{1}{}{ } & \multicolumn{2}{|c|}{3$\nu$ }& \multicolumn{2}{|c|}{3+1 }\\
\cline{2-3}\cline{4-5}
\cline{2-3}\cline{4-5}
 \multicolumn{1}{}{ } &\multicolumn{1}{|c|}{NH} & \multicolumn{1}{c|}{IH} & \multicolumn{1}{|c|}{NH} & \multicolumn{1}{c|}{IH} \\
\hline
\multicolumn{1}{||c||}{$c_{ee}$} & $|U_{e3}|^2$=0.02  & $|U_{e2}|^2$=0.30  & $|U_{e4}|^2=0.02$& $|U_{e4}|^2=0.02$ \\
\hline
\multicolumn{1}{||c||}{$c_{xe}$}  &$1-|U_{e3}|^2=0.98$& $1-|U_{e2}|^2=0.70$  & $|U_{e1}|^2+ |U_{e2}|^2=0.96$  & $|U_{e1}|^2+ |U_{e3}|^2=0.69$\\
\hline
\multicolumn{1}{||c||}{$c_{se}$}   & -&- &$|U_{e3}|^2=0.02$ & $|U_{e2}|^2=0.29$ \\
\hline
\end{tabular}
\label{tabI}
\end{center}
\end{table}

The $\nu_e$ flux at the Earth would share the same flavor composition computed above at the exit of the SN, but for the {\it different kinematics} characterizing the propagation of neutrinos of different masses. Since the original $\nu_e$ flux completely converts to $\nu_4$, the part of spectrum proportional to $F_{\nu_e}^0$ gets delayed and broadened in time with respect to the other components, where the other components correspond to the $e$-flavor projections of the ``light'' states. So, when making explicit the time and energy-dependence of the fluxes, apart from the geometrical factor $\propto (4\pi D^{2})^{-1}$, the flux at the Earth writes (assuming the 3 lightest states have vanishingly small masses and setting $F^0_{\nu_s}=0$)
\begin{equation}
F_{{\nu}_e}(E_\nu,t)\approx |U_{e4}|^2F^0_{\nu_e}\left(E_\nu,t-\frac{D}{2c}\left(\frac{m_4}{E_\nu}\right)^2\right)+(1-|U_{e4}|^2-|U_{ei}|^2) F^0_{\nu_x}(E_\nu,t)~,\label{approxres}
\end{equation}
with $i=2,3$ for the inverted or normal hierarchies, respectively. Obviously the delay in the component of the $\nu_e$ flux proportional to $F^0_{\nu_e}$ depends on the distance of supernova, $D$, the mass of the heaviest state and the neutrino energy, such that
\begin{equation}
\frac{D}{2c} \left( \frac{m_4}{E_\nu} \right)^2 = 5.15\; {\rm ms} \left( \frac{D}{10\,{\rm kpc}} \right) \left( \frac{10 \, {\rm MeV}}{E_\nu} \right)^2 \left( \frac{m_4}{1\,{\rm eV}} \right)^2\,.
\end{equation}

\subsection{Phenomenological Considerations on SN $\nu_e$ Flux at Earth}\label{sec:flux}

\begin{figure}[t!]
\centering
\subfloat[]{
 \includegraphics[trim= 0mm 0mm 0mm 
0mm,clip,width=0.5\textwidth]{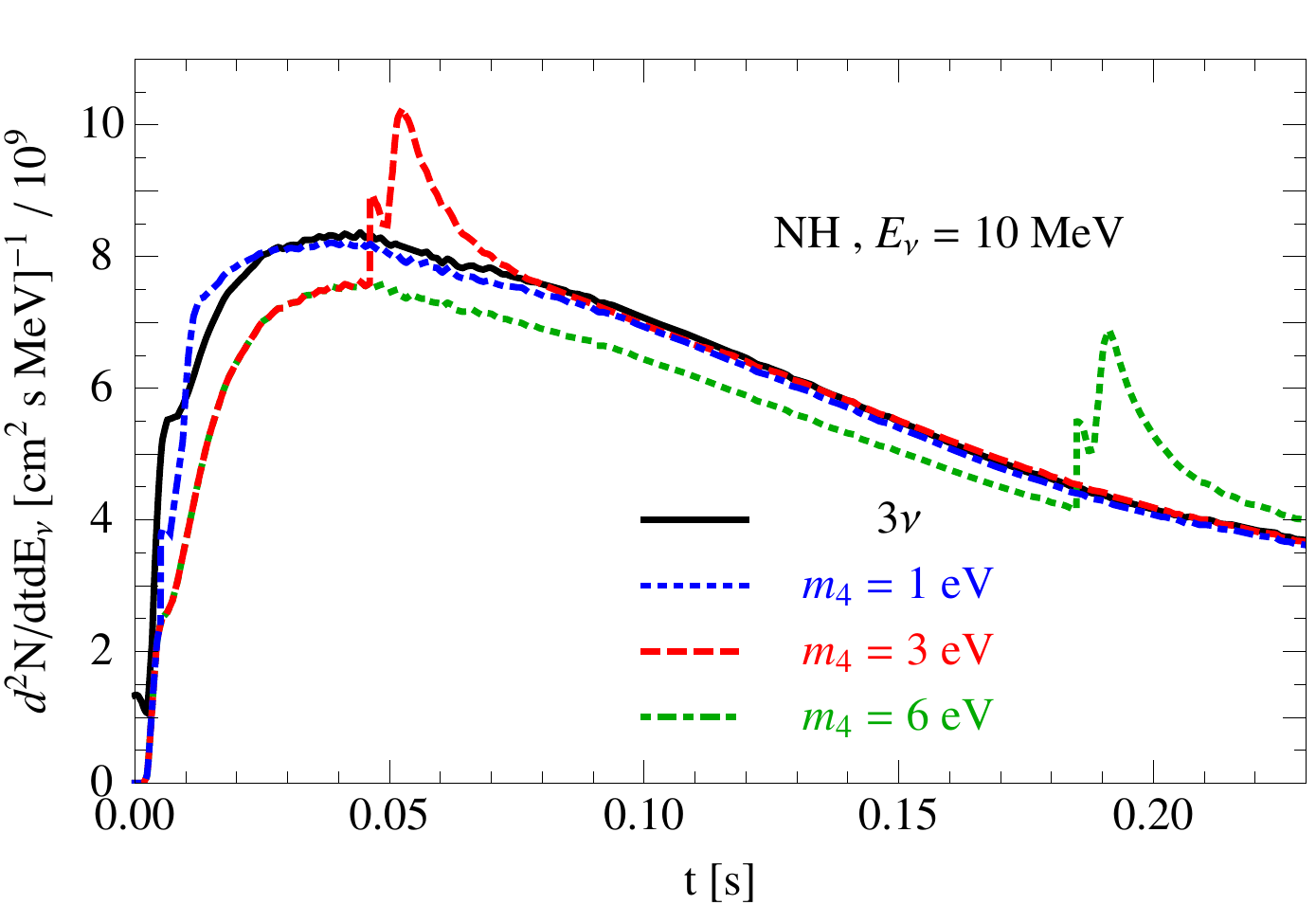}
  \label{fig:flux-NH}
}
\subfloat[]{
 \includegraphics[trim= 0mm 0mm 0mm 
0mm,clip,width=0.5\textwidth]{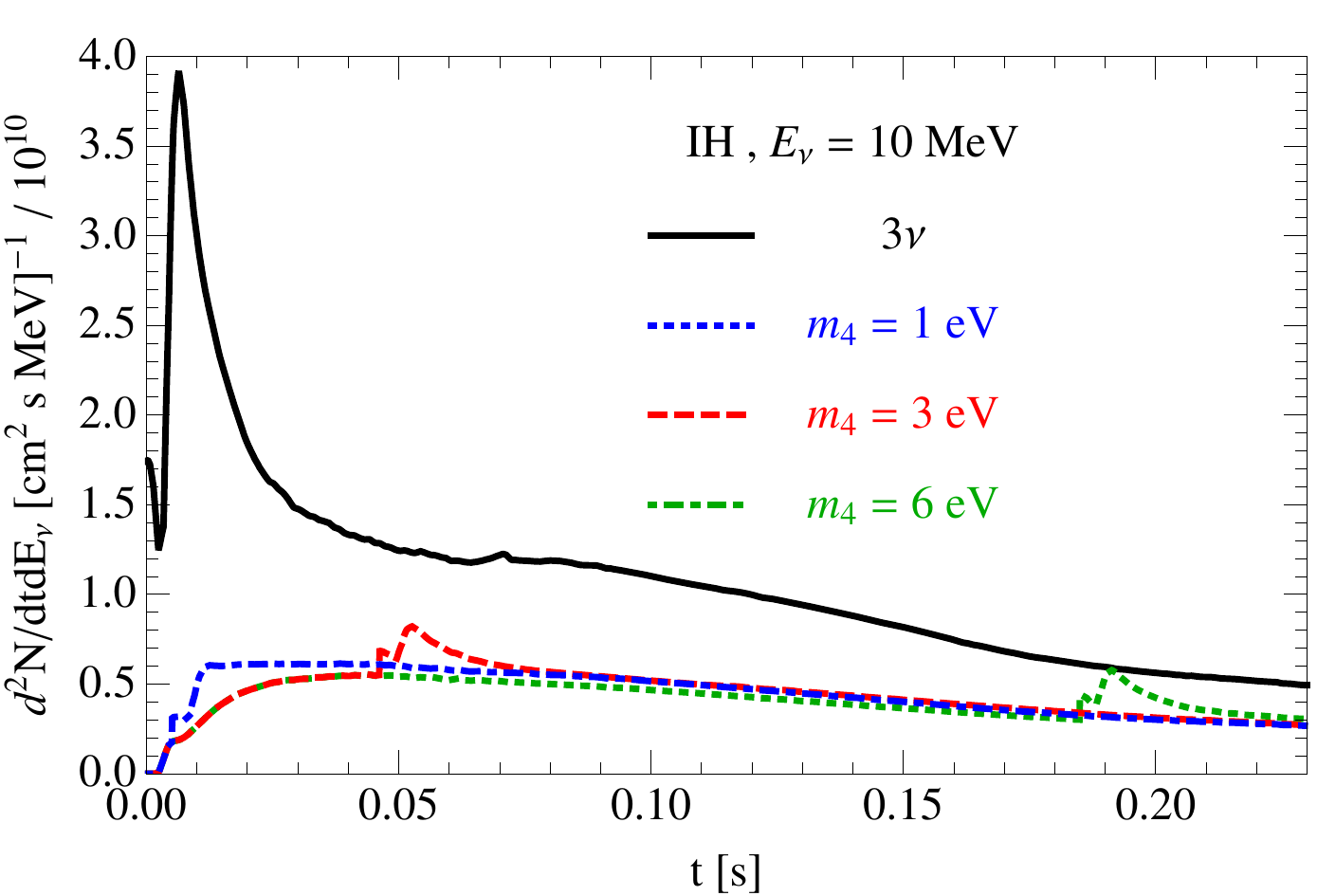}
  \label{fig:flux-IH}
}
\quad
\subfloat[]{
 \includegraphics[trim= 0mm 0mm 0mm 
0mm,clip,width=0.5\textwidth]{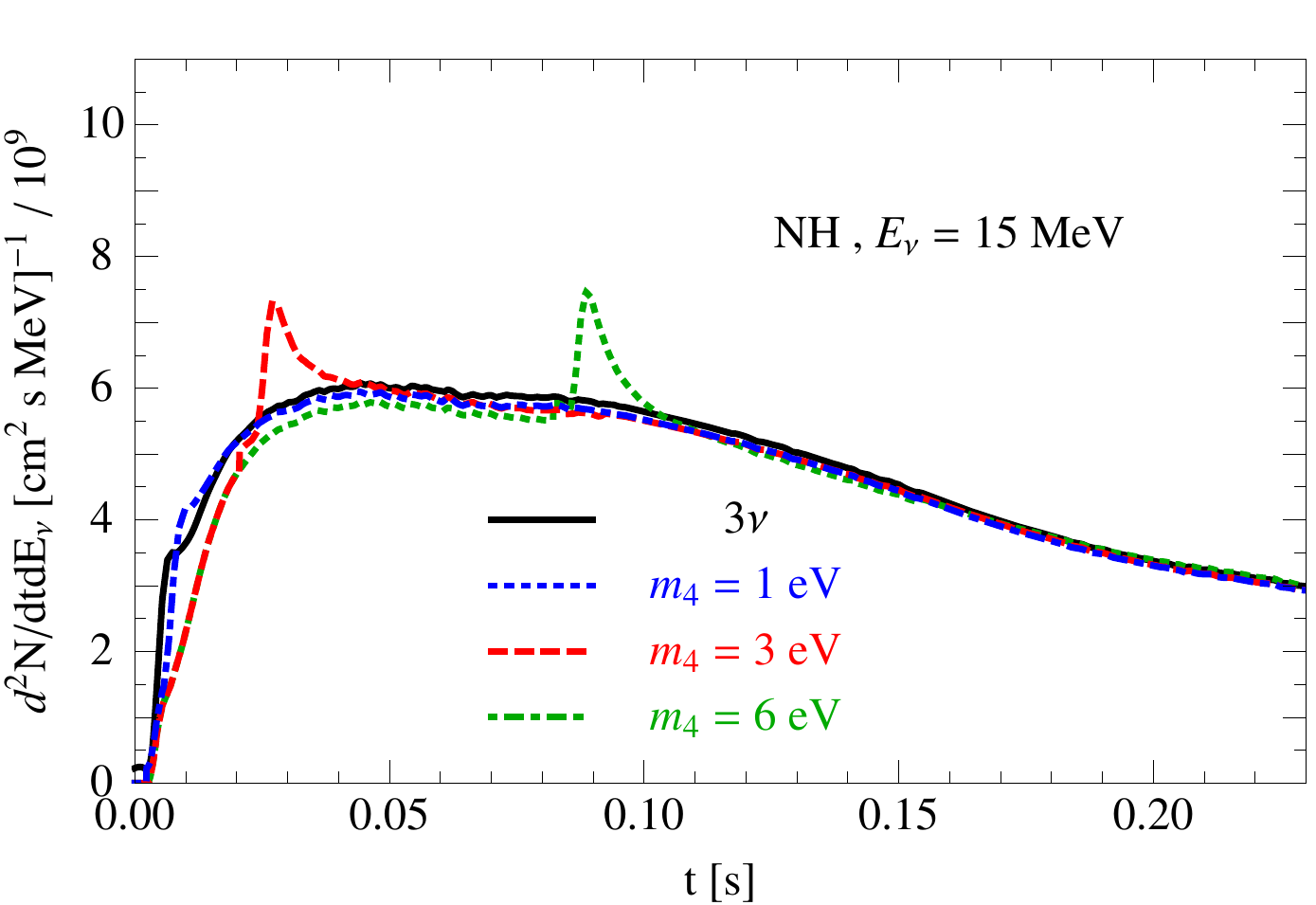}
  \label{fig:flux-NH1}
}
\subfloat[]{
 \includegraphics[trim= 0mm 0mm 0mm 
0mm,clip,width=0.5\textwidth]{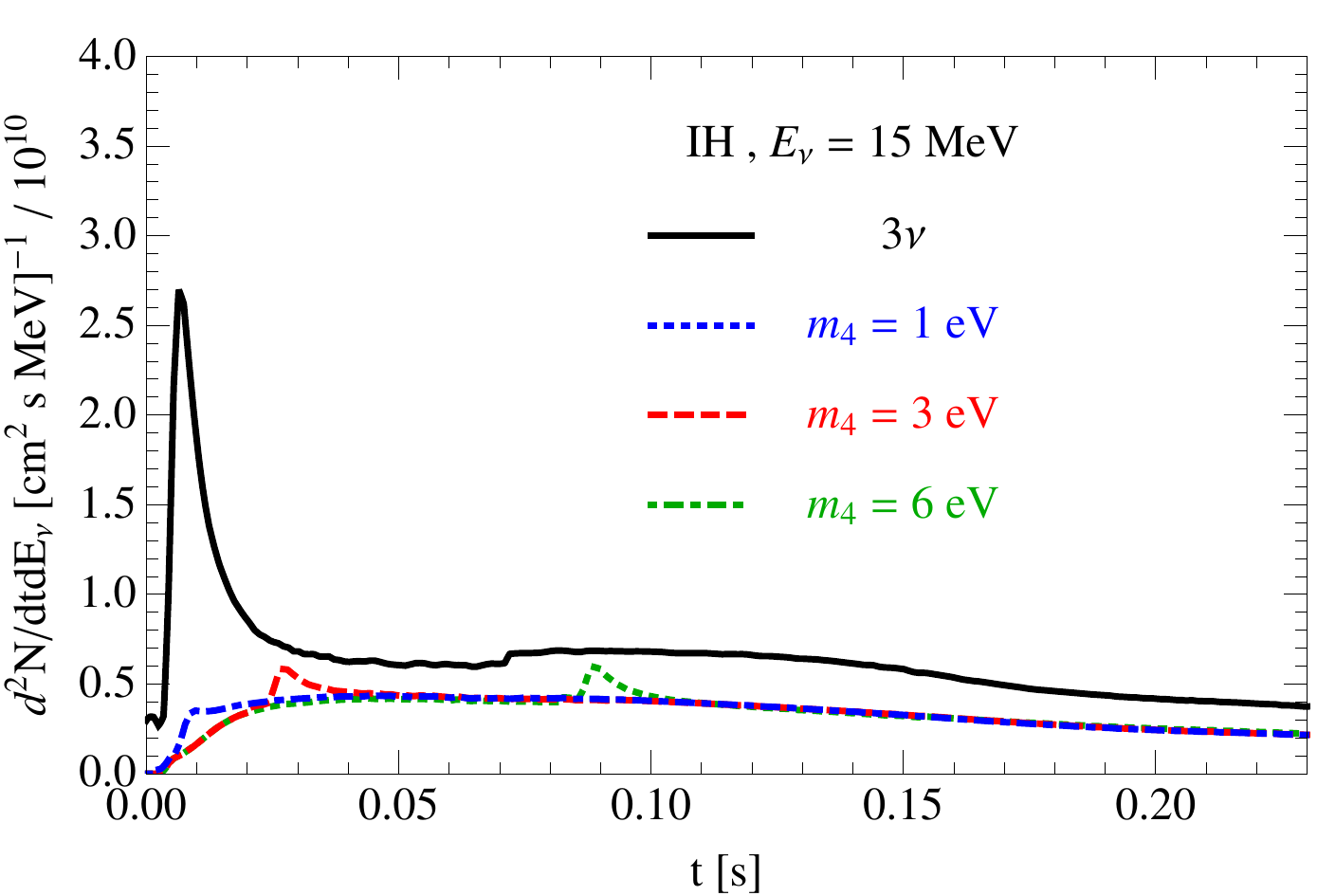}
  \label{fig:flux-IH2}
}
\caption{\label{fig:flux}The flux $F_{\nu_e}=d^2 N_\nu/dtdE_\nu$ at Earth for: (a) NH and $E_\nu=10$~MeV; (b) IH and $E_\nu=10$~MeV; (c) NH and $E_\nu=15$~MeV; (d) IH and $E_\nu=15$~MeV. In this figure we assume SN distance $D=10$~kpc and $(\theta_{14},\theta_{24},\theta_{34})=(8.7^\circ,0,0)$.}
\end{figure}

\begin{figure}[!ht]
\centering
\subfloat[$3\nu$ framework, NH]{
\includegraphics[trim= 0mm 0mm 0mm 
100mm,clip,width=0.3\textwidth]{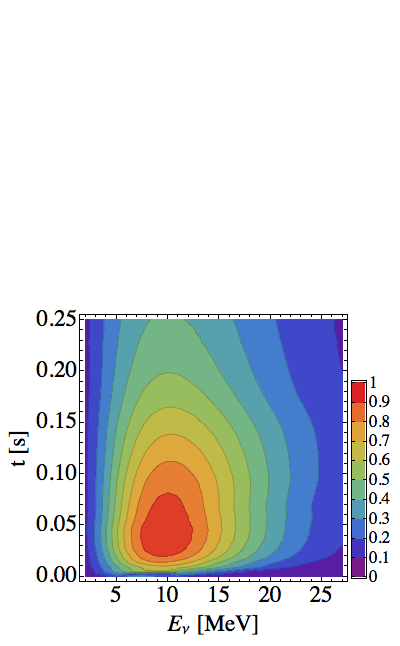}
  \label{fig:stdNH}
}
\subfloat[$3\nu$ framework, IH]{
 \includegraphics[trim= 0mm 0mm 0mm 
100mm,clip,width=0.3\textwidth]{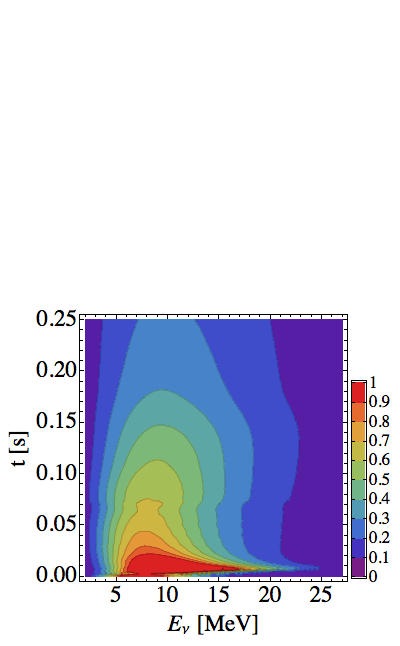}
  \label{fig:stdIH}
}
\quad
\subfloat[$3+1$ model, NH, $m_4=1$~eV]{
 \includegraphics[trim= 0mm 0mm 0mm 
100mm,clip,width=0.3\textwidth]{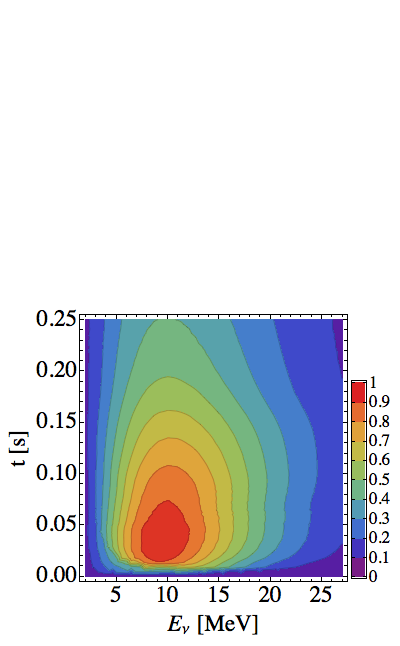}
  \label{fig:m1evNH-2d}
}
\subfloat[$3+1$ model, IH, $m_4=1$~eV]{
 \includegraphics[trim= 0mm 0mm 0mm 
100mm,clip,width=0.3\textwidth]{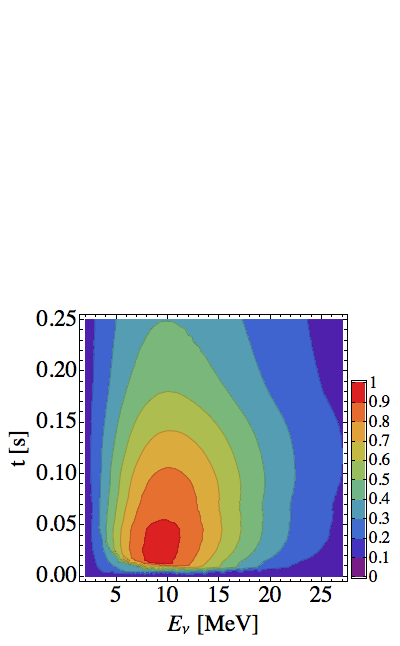}
  \label{fig:m1evIH-2d}
}
\quad
\subfloat[$3+1$ model, NH, $m_4=3$~eV]{
\includegraphics[trim= 0mm 0mm 0mm 
100mm,clip,width=0.3\textwidth]{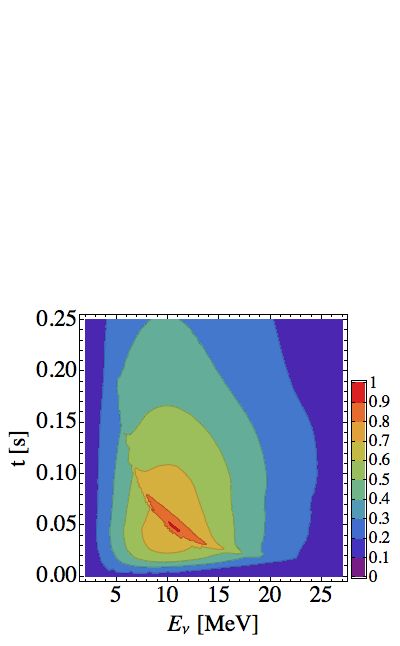}
  \label{fig:m3evNH-2d}
}
\subfloat[$3+1$ model, IH, $m_4=3$~eV]{
 \includegraphics[trim= 0mm 0mm 0mm 
100mm,clip,width=0.3\textwidth]{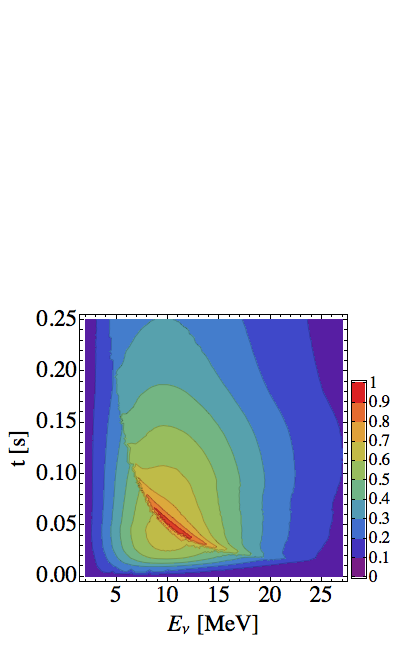}
  \label{fig:m3evIH-2d}
}
\quad
\subfloat[$3+1$ model, NH, $m_4=6$~eV]{
 \includegraphics[trim= 0mm 0mm 0mm 
100mm,clip,width=0.3\textwidth]{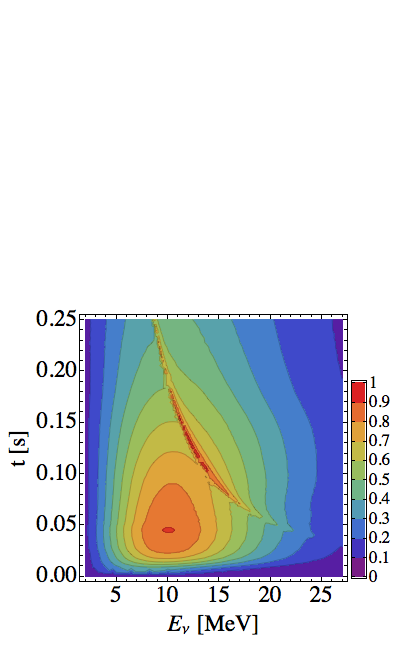}
  \label{fig:m6evNH-2d}
}
\subfloat[$3+1$ model, IH, $m_4=6$~eV]{
 \includegraphics[trim= 0mm 0mm 0mm 
100mm,clip,width=0.3\textwidth]{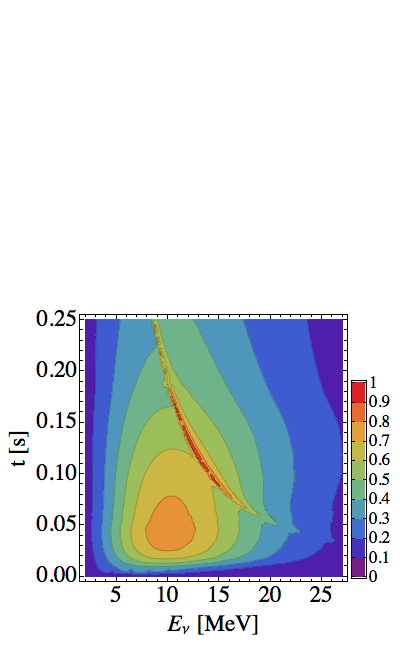}
  \label{fig:m6evIH-2d}
}\caption{\label{fig:plot-2d}The contour plot of flux $F_{\nu_e}=d^2 N_\nu/dtdE_\nu$ at Earth for NH (left column panels) and IH (right column panels). From top to bottom rows: the 3$\nu$ case, the $3+1$ model with $m_4=1$~eV, $m_4=3$~eV and $m_4=6$~eV, respectively. In all the panels we assume $(\theta_{14},\theta_{24},\theta_{34})=(8.7^\circ,0,0)$. In all the panels the flux is normalized to the maximum value.}
\end{figure}

To illustrate the effect of sterile neutrinos on SN $\nu_e$ flux, discussed in Eq.~(\ref{approxres}), we plot in Figure~\ref{fig:flux} the $F_{\nu_e}$ at Earth as function of time for NH (left panels) and IH (right panels) for the $3\nu$ framework and for the $3+1$ model with $m_4=1,3$ and $6$~eV. In the top (bottom) panels we assume $E_\nu=10$~MeV (15~MeV). For the mixing angles in Figure~\ref{fig:flux} we take $\theta_{14}=8.7^\circ$ and $\theta_{24}=\theta_{34}=0$ (although the plots are the same for nonzero $\theta_{24}$ and $\theta_{34}$). For the NH case, the differences with respect to the standard 3$\nu$ case are relatively moderate. Most notably, the existence of the sterile neutrino leads to the appearance of a small peak (originated from neutronization burst) whose height is proportional to $|U_{e4}|^2$ and whose delay with respect to the bounce time is proportional to $m_4^2$ (assuming a fixed value of $E_\nu$ and SN distance $D$). For the IH, however, the modification is huge: the expected neutronization burst in $3\nu$ disappears, as we anticipated. On  top of that, a smaller peak reappears at later times, with the same features discussed for NH. As we mentioned, the distortion of $\nu_e$ flux due to kinematical effects depends on both $m_4$ and $E_\nu$ (see Eq.~(\ref{approxres})) for a fixed distance of SN. To illustrate this dependence, in Figure~\ref{fig:plot-2d} we show the contour plots of $F_{\nu_e}(E_\nu,t)$. In Figure~\ref{fig:plot-2d} the left (right) panels are for NH (IH) and, from top to bottom, panels correspond to $3\nu$ framework and $3+1$ model with $m_4=1,3$ and $6$~eV. In all the panels for $3+1$ model we assume $(\theta_{14},\theta_{24},\theta_{34})=(8.7^\circ,0,0)$. Clearly the delayed component structure $\propto E_{\nu}^{-2}$ can be seen. The structure of delayed peak is the same for NH or IH. Note that for masses smaller than $1\,$eV the picture would look very similar to the $1\,$eV case.

In summary, in the $3\nu$ framework the observation of $\nu_e$ burst strongly points to IH for neutrino mass scheme; while this conclusion can completely change in the presence of a sterile neutrino. On one hand, we can conclude that the observation of the expected burst would not only indicate the IH of the active neutrinos, but also {\it exclude the presence of sterile neutrinos with mass-mixing parameters possibly unaccessible to the other terrestrial experiments.} On the other hand, in the $3+1$ model  the {\it non-observation} of the burst does not allow any immediate conclusion on the active neutrino mass hierarchy. In particular for small $m_4$ and small $\theta_{14}$, the time-energy profiles of the second row in Figure~\ref{fig:plot-2d} are not only quite similar to each other, but also to the $3\nu$ NH case of the top row. Better diagnostics in this case requires further information, either from external input or from the SN signal itself. For example, if at the time of the Galactic SN detection one knew that active neutrinos have IH, the absence of a detectable burst (provided that one has a sufficiently sensitive detector, of course) could be interpreted as a signature of a sterile neutrino. By the way, this signature is present also for mixing angles too small to be detected in the terrestrial experiments, which is an interesting complementarity of this astroparticle detection channel with respect to terrestrial probes. 

Needless to say, independently of the mass hierarchy, if a delayed small peak were detected one could constrain the sterile neutrino mass-mixing parameters and also identify that this mechanism is at play. Note that the neutronization burst has been discussed in the past as a way to constrain active neutrino masses, see e.g.~\cite{Arnaud:2001gt} for an early proposal and \cite{Pagliaroli:2010ik} for a more recent discussion in the context of different neutrino mass determination methods. One of the main difficulties in SN neutrino mass determination methods is due to the fact that current cosmological constraints push towards a relatively low neutrino mass scale, say of the order of ${\cal O}$(0.1)~eV, for which the above mentioned kinematical effects are negligible. The delayed peak effect linked to sterile neutrinos stressed here presents however different types of challenges: on the one hand, the delay can be significantly more important and ease its detection. On the other hand, it is typically a small effect. Although the experimental verification of the suppressed peak would be challenging, the reward would be also great; hence we foresee further (detector-specific) studies in the future.

\section{SN $\overline{\nu}_e$ flux in $3+1$ scenario\label{phenonuebar}}

In this section we briefly discuss the antineutrino sector, since existing detectors are mostly sensitive to $\overline{\nu}_e$. For the flux of $\overline{\nu}_e$ at Earth\footnote{Here we ignore the Earth matter effect. Its detectability in a forthcoming Galactic SN event has been re-evaluated recently in~\cite{Borriello:2012zc} in the light of recent simulation results and found quite dim, in any case.} we can write
\begin{eqnarray}\label{eq:nuebarflux}
F_{\overline{\nu}_e} &=& \bar{c}_{ee} F^0_{\overline{\nu}_e} + \bar{c}_{xe} F^0_{\overline{\nu}_x} + \bar{c}_{se} F^0_{\overline{\nu}_s}~.
\end{eqnarray}
The expressions for coefficients $(\bar{c}_{ee},\bar{c}_{xe})$ in $3\nu$ framework and their numerical values (for best-fit values of mixing angles) are shown in the first column of Table~\ref{tabII}. In the second and third columns of Table~\ref{tabII} the expressions for $(\bar{c}_{ee},\bar{c}_{xe},\bar{c}_{se})$ coefficients in $3+1$ model for the cases of vanishing and non-vanishing $\{\theta_{24},\theta_{34}\}$ are reported, respectively. The analytical results reported Table~\ref{tabII} (which can be derived straightforwardly from the level crossing scheme of antineutrinos and whose details are reported in Appendix~\ref{FluxEarth}) have been again cross-checked numerically and found in excellent agreement; for numerical errors, similar considerations to the ones for neutrinos in Table~\ref{tabI} apply. In the case of $\theta_{24}=\theta_{34}=0$, if we neglect differences at the few-percent level, the presence of the sterile-state does not imply appreciable differences in the outgoing $\overline{\nu}_e$ flux composition. This has been noted before, see e.g.~\cite{Tamborra:2011is}, and crucially depends on the fact that we assumed $U_{e4}$ is the only non-vanishing mixing element in the fourth column of mixing matrix.

\begin{table}[!htb]
\caption{The coefficients in Eq.~(\ref{eq:nuebarflux}) in $3\nu$ and $3+1$ models. For the numerical values we set $\theta_{14}=8.7^\circ$ in the second column and $(\theta_{24},\theta_{34})= (9.8^\circ,0)$ in the third column.}\begin{center}
\begin{tabular}{|c|c|c|c|c|c|c|}
\cline{2-3}\cline{4-5}\cline{6-7}
\cline{2-3}\cline{4-5}\cline{6-7}
\multicolumn{1}{}{ } & \multicolumn{2}{|c|}{3$\nu$}& \multicolumn{2}{|c|}{$3+1$, $\theta_{14}\neq0$ and $\theta_{24}=\theta_{34}=0$ }& \multicolumn{2}{|c|}{3+1, $\theta_{24}\neq0$ and/or $\theta_{34}\neq0$ }\\
\cline{2-3}\cline{4-5}\cline{6-7}
\cline{2-3}\cline{4-5}\cline{6-7}
 \multicolumn{1}{}{ } & \multicolumn{1}{|c|}{NH} & \multicolumn{1}{c|}{IH}  & \multicolumn{1}{|c|}{NH } & \multicolumn{1}{c|}{IH }& \multicolumn{1}{|c|}{NH } & \multicolumn{1}{c|}{IH} \\
\hline
\multicolumn{1}{||c||}{$\bar{c}_{ee}$} & $|U_{e1}|^2$=0.68 &  $|U_{e3}|^2$=0.02 & $|U_{e1}|^2=0.66$ &$|U_{e3}|^2 = 0.02$ &$|U_{e1}|^2=0.66$ & $|U_{e3}|^2 = 0.02$  \\
\hline
\multicolumn{1}{||c||}{$\bar{c}_{xe}$}  &$1-|U_{e1}|^2$=0.32 & $1-|U_{e3}|^2=0.98$  & $|U_{e2}|^2 + |U_{e3}|^2 = 0.31$ & $|U_{e1}|^2 + |U_{e2}|^2 = 0.96$&  $|U_{e3}|^2 + |U_{e4}|^2 = 0.05 $ & $|U_{e2}|^2 + |U_{e4}|^2 = 0.32$ \\
\hline
\multicolumn{1}{||c||}{$\bar{c}_{se}$}  & -&-  &   $|U_{e4}|^2 = 0.02$ &  $|U_{e4}|^2=0.02$  & $|U_{e2}|^2 = 0.29$ & $|U_{e1}|^2 = 0.66$ \\
\hline
\end{tabular}
\label{tabII}
\end{center}
\end{table}

However, this conclusion is not robust against non-vanishing 2-4 and 3-4 mixings: even small nonzero values of $\theta_{24}$ and/or $\theta_{34}$ lead to resonant conversions $\overline{\nu}_\mu \to \overline{\nu}_4$ and $\overline{\nu}_\tau \to \overline{\nu}_4$, respectively, with consequent alteration in antineutrino fluxes. In particular, the current upper limit on $\theta_{34}$ or $U_{\tau4}$ is so poor ($|U_{\tau4}|^2 \lesssim 0.2$ at 90\% C.L.)~\cite{Kopp:2013vaa} that there is ample margin for a sizable alteration of the $\overline{\nu}_e$ SN flux via a finite $\nu_\tau-\nu_s$ mixing. For a more concrete benchmark case, we can assume $\theta_{24}= 9.8^\circ$ inspired by the best-fit values of the global analyses in~\cite{Kopp:2013vaa,Giunti:2013aea,Giunti:2011cp}. In this case, the coefficients in Eq.~(\ref{eq:nuebarflux}) are given in the last column of Table~\ref{tabII}. It is clear that the $\overline{\nu}_e$ flux composition is now appreciably different, due to changes in $\bar{c}_{xe}$, which quantifies the $\overline{\nu}_x \to \overline{\nu}_e$ oscillation probability changes due to the resonance in $\overline{\nu}_\mu - \overline{\nu}_s$ channel (since we assumed only $\theta_{24}\neq0$). For the NH case, $\bar{c}_{xe}$ drops by one order of magnitude: this implies that the final $\overline{\nu}_e$ flux loses almost completely the contribution from the initial $\overline{\nu}_x$ flux (the initial $\overline{\nu}_x$ state mostly converted into a sterile state). The consequences are perhaps not dramatic, since two thirds of the flux come from the initial $\overline{\nu}_e$, roughly like in the standard $3\nu$ scenario. Yet, differences of the order of 30\% are expected assuming comparable initial fluxes and may lead to observable consequences. In the IH case, however, the value of $\bar{c}_{ee}=|U_{e3}|^2$ in the standard 3$\nu$ case is very small: in the standard scenario most  of the observable $\overline{\nu}_e$ flux comes from the initial $\overline{\nu}_x$ one. But now in presence of $\nu_s$ the coefficient $\bar{c}_{xe}$ is reduced by a {\it factor of three}! A major alteration in the flux is expected, with consequences for the time-dependent luminosity profile in detectors such as IceCube~\cite{Halzen:1994xe} or the number, energy and time distribution of events in a Water Cherenkov detector. A factor of three is well above the flux differences due to different progenitors (see e.g. Fig.~1 in~\cite{Serpico:2011ir}) and even the overall number of events may  already constitute an interesting diagnostic channel, especially if the progenitor type and distance could be identified. We plan to treat the observational consequences of these effects in more detail (and in a detector-dependent way) in a forthcoming publication.

\begin{figure}[t!]
\centering
\subfloat[]{
 \includegraphics[trim= 0mm 0mm 0mm 
0mm,clip,width=0.5\textwidth]{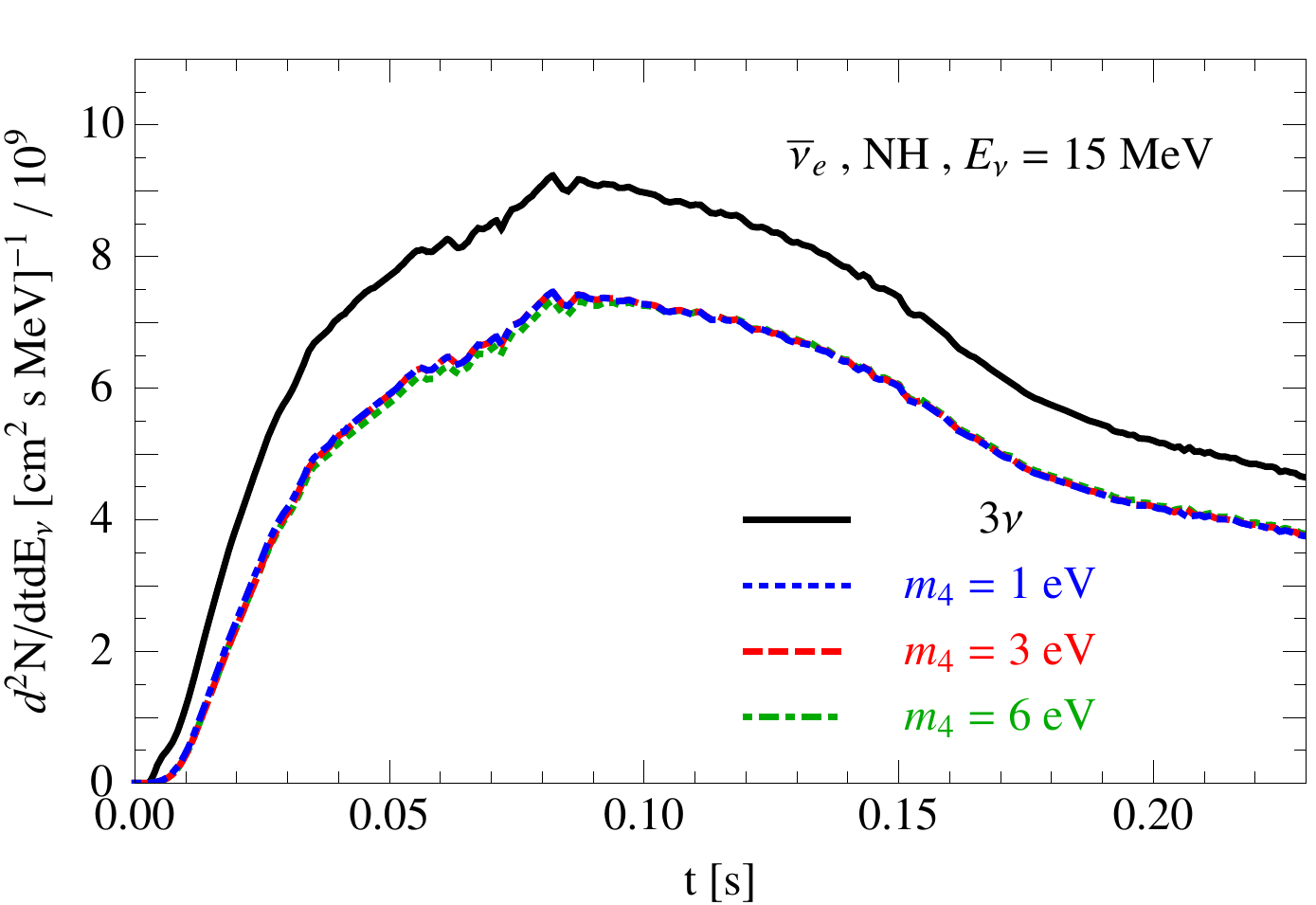}
  \label{fig:flux-NH1}
}
\subfloat[]{
 \includegraphics[trim= 0mm 0mm 0mm 
0mm,clip,width=0.5\textwidth]{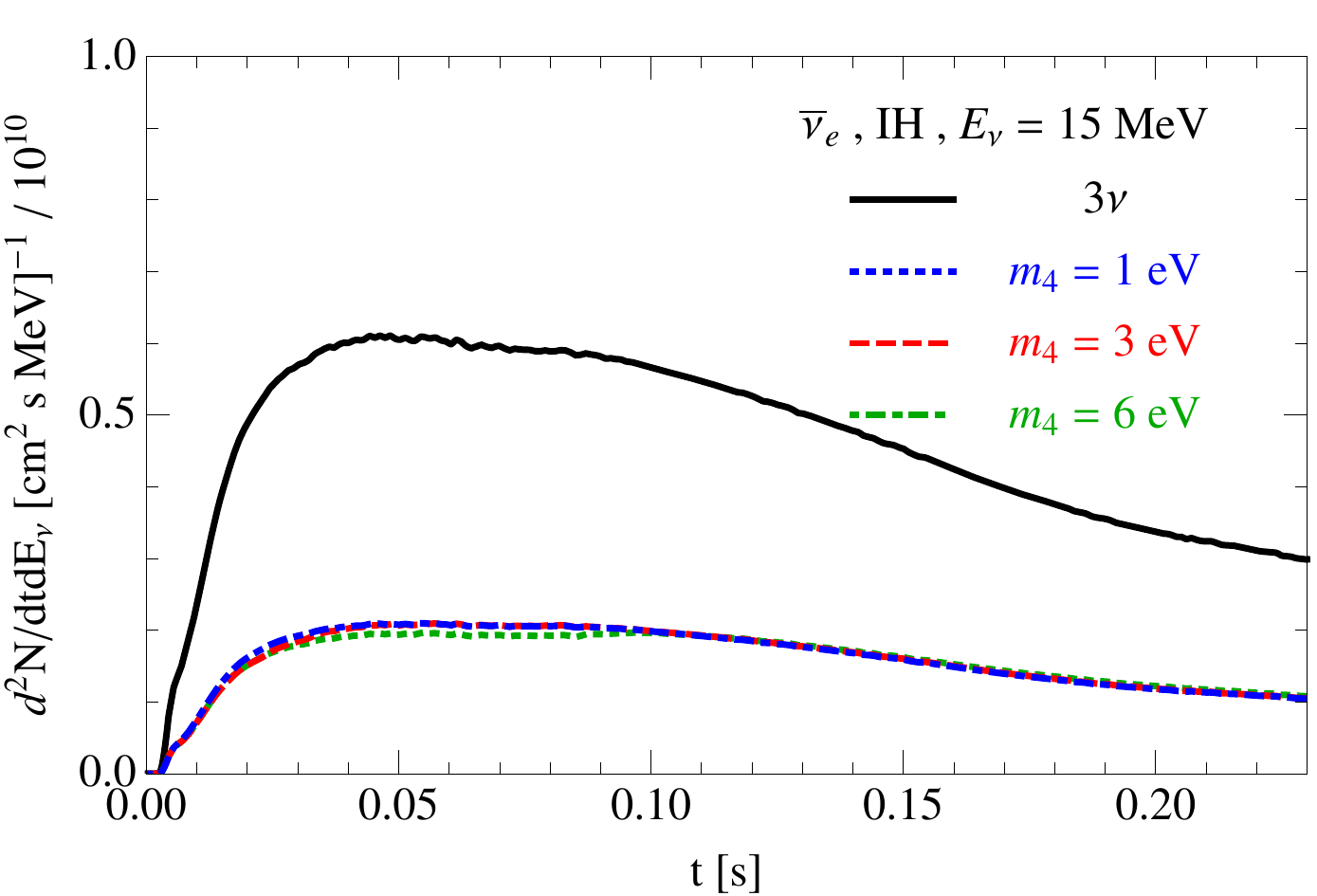}
  \label{fig:flux-IH2}
}
\caption{\label{fig:fluxnuebar}The flux $F_{\overline{\nu}_e}=d^2 N_\nu/dtdE_\nu$ at Earth for: (a) NH and $E_\nu=15$~MeV and (b) IH and $E_\nu=15$~MeV. In this figure we assume SN distance $D=10$~kpc and $(\theta_{14},\theta_{24},\theta_{34})=(8.7^\circ,9.8^\circ,0)$.}
\end{figure}

As we mentioned, in the case of $\theta_{14}\neq0$ and $\theta_{24}=\theta_{34}=0$, the $\overline{\nu}_e$ flux composition in $3\nu$ and $3+1$ model are similar. Also, since there is no resonance conversion for antineutrinos in this case, none of the components will be delayed. But, in the case of $\theta_{24}\neq0$ and/or $\theta_{34}\neq0$, since the initial $\overline{\nu}_x$ almost completely converts to $\overline{\nu}_4$, the contribution of $F_{\overline{\nu}_x}^0$ to the $\overline{\nu}_e$ flux will be delayed. In fact, when $\theta_{24}\neq0$ and $\theta_{34}=0$, during the propagation in SN all the initial $\overline{\nu}_\mu$ converts to $\overline{\nu}_4$ while $\overline{\nu}_\tau$ goes to $\overline{\nu}_3$ ($\overline{\nu}_2$) for NH (IH). The conversion pattern for the case $\theta_{24}=0$ and $\theta_{34}\neq0$ is the opposite; {\it i.e.}, $\overline{\nu}_\tau$ converts to $\overline{\nu}_4$ while $\overline{\nu}_\mu$ goes to $\overline{\nu}_3$ ($\overline{\nu}_2$) for NH (IH). When both $\theta_{24}\neq0$ and $\theta_{34}\neq0$, although $\overline{\nu}_\mu$ and $\overline{\nu}_\tau$ convert to both $\overline{\nu}_4$ and $\overline{\nu}_3$ ($\overline{\nu}_2$) for NH (IH), since the initial flux of $\overline{\nu}_\mu$ and $\overline{\nu}_\tau$ are the same at production region, effectively one $F^0_{\overline{\nu}_x}$ ($x=\mu$ or $\tau$) converts to $F_{\overline{\nu}_4}$ at the surface of SN. Taking into account all these subtleties, the kinematical effect in the presence of sterile neutrino with $\theta_{24}\neq0$ and/or $\theta_{34}\neq0$ on the $\overline{\nu}_e$ flux can be written as
 \begin{equation}
F_{\overline{\nu}_e}(E_\nu,t)\approx \bar{c}_{ee} F^0_{\overline{\nu}_e}\left(E_\nu,t\right)+|U_{ei}|^2F^0_{\overline{\nu}_x}\left(E_\nu,t\right)+|U_{e4}|^2F^0_{\overline{\nu}_x}\left(E_\nu,t -\frac{D}{2c}\left(\frac{m_4}{E_\nu}\right)^2\right)~,\label{approxresnubar}
\end{equation}
where $i=2,3$ for IH and NH, respectively; and the coefficient $\bar{c}_{ee}$ is given in the third column of Table~\ref{tabII}. To illustrate the impact of sterile neutrino of $\overline{\nu}_e$ flux, in Figure~\ref{fig:fluxnuebar} we show $F_{\overline{\nu}_e}$ for both NH and IH for the energy $E_\nu=15$~MeV. In this figure we assume $(\theta_{14},\theta_{24},\theta_{34})=(8.7^\circ,9.8^\circ,0)$. As we discussed, for NH the effect is a moderate reduction in flux; while for the IH a more significant reduction can be seen.

\begin{figure}[!ht]
\centering
\subfloat[$\overline{\nu}_e$, $3\nu$ framework, NH]{
\includegraphics[trim= 0mm 0mm 0mm 
100mm,clip,width=0.3\textwidth]{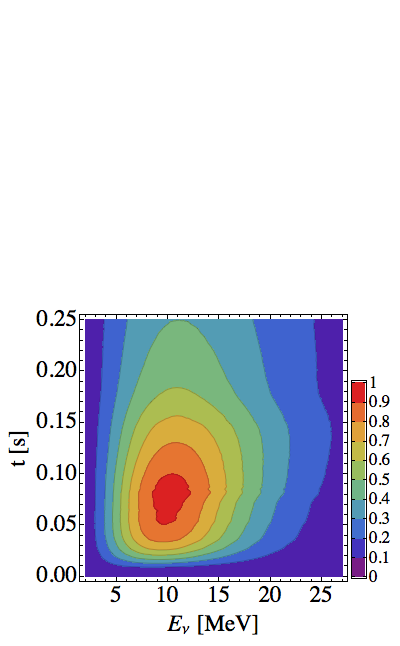}
  \label{fig:stdNHnubar}
}
\subfloat[$\overline{\nu}_e$, $3\nu$ framework, IH]{
 \includegraphics[trim= 0mm 0mm 0mm 
100mm,clip,width=0.3\textwidth]{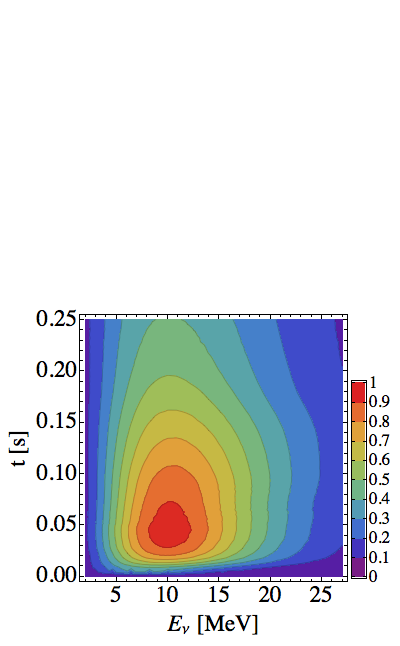}
  \label{fig:stdIHnubar}
}
\quad
\subfloat[$\overline{\nu}_e$, $3+1$ model, NH, $m_4=1$~eV]{
 \includegraphics[trim= 0mm 0mm 0mm 
100mm,clip,width=0.3\textwidth]{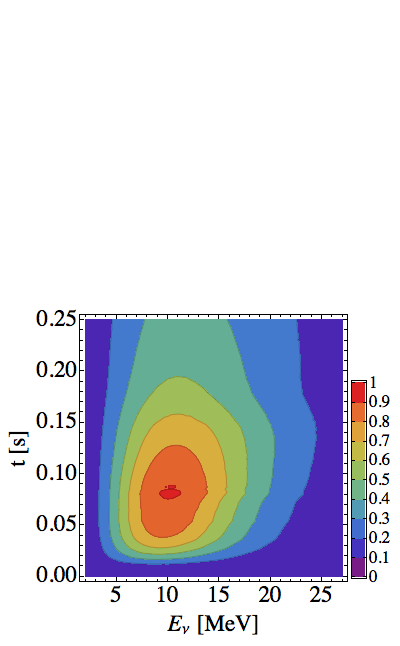}
  \label{fig:m1evNH-2dnubar}
}
\subfloat[$\overline{\nu}_e$, $3+1$ model, IH, $m_4=1$~eV]{
 \includegraphics[trim= 0mm 0mm 0mm 
100mm,clip,width=0.3\textwidth]{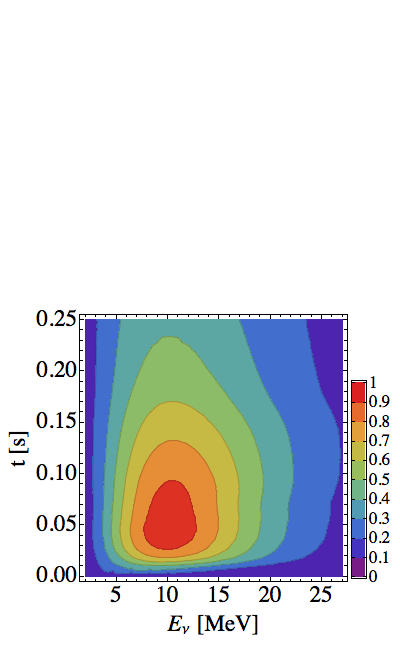}
  \label{fig:m1evIH-2dnubar}
}
\quad
\subfloat[$\overline{\nu}_e$, $3+1$ model, NH, $m_4=3$~eV]{
\includegraphics[trim= 0mm 0mm 0mm 
100mm,clip,width=0.3\textwidth]{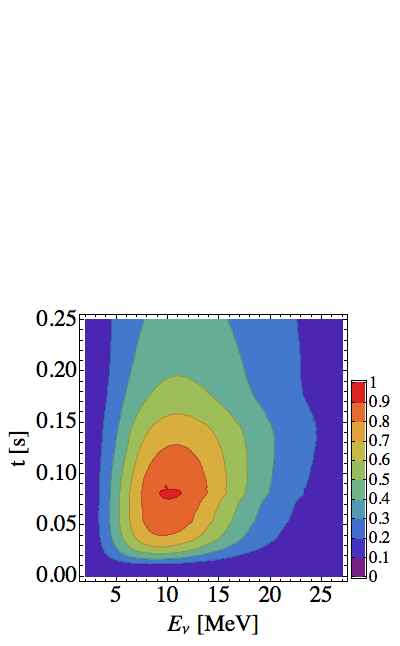}
  \label{fig:m3evNH-2dnubar}
}
\subfloat[$\overline{\nu}_e$, $3+1$ model, IH, $m_4=3$~eV]{
 \includegraphics[trim= 0mm 0mm 0mm 
100mm,clip,width=0.3\textwidth]{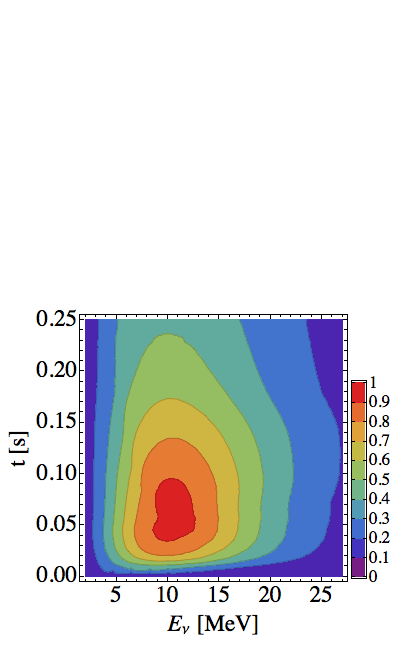}
  \label{fig:m3evIH-2dnubar}
}
\quad
\subfloat[$\overline{\nu}_e$, $3+1$ model, NH, $m_4=6$~eV]{
\includegraphics[trim= 0mm 0mm 0mm 
100mm,clip,width=0.3\textwidth]{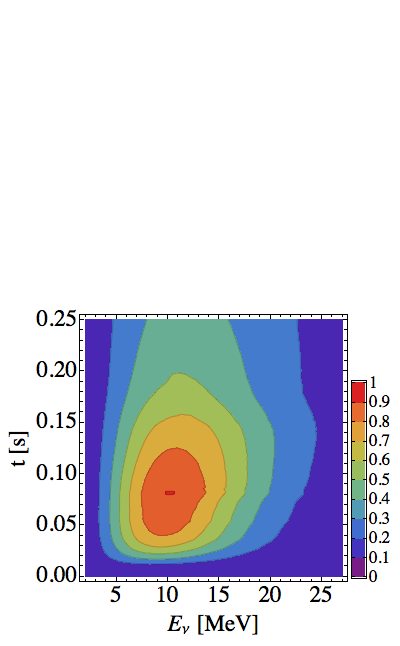}
  \label{fig:m3evNH-2dnubar}
}
\subfloat[$\overline{\nu}_e$, $3+1$ model, IH, $m_4=6$~eV]{
 \includegraphics[trim= 0mm 0mm 0mm 
100mm,clip,width=0.3\textwidth]{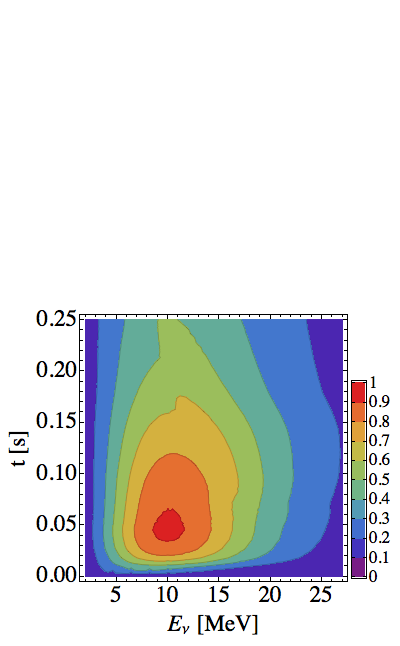}
  \label{fig:m3evIH-2dnubar}
}
\caption{\label{fig:plot-2dnubar}The contour plot of flux $F_{\overline{\nu}_e}=d^2 N_\nu/dtdE_\nu$ at Earth for NH (left column panels) and IH (right column panels). From top to bottom rows: the 3$\nu$ case, the $3+1$ model with $m_4=1$~eV,  $m_4=3$~eV and $m_4=6$~eV, respectively. In all the panels we assume $(\theta_{14},\theta_{24},\theta_{34})=(8.7^\circ,9.8^\circ,0)$. In all the panels the flux is normalized to the maximum value.}
\end{figure}

Also, in Figure~\ref{fig:plot-2dnubar} we show the contour plots of $F_{\overline{\nu}_e}(E_\nu,t)$ for $3\nu$ and $3+1$ model with $m_4=1$, $3$ and $6$~eV, for both NH and IH. As can be seen, in the NH case the effect is almost negligible, while in IH case since the main contribution to $F_{\overline{\nu}_e}$ is from $F_{\overline{\nu}_x}$, moderate distortion are more notable.

\section{Discussion and conclusions}\label{discussion}

The next Galactic supernova explosion and the observation of its neutrino flux in the existing and forthcoming experiments at Earth will provide a unique opportunity to study both the explosion mechanism and neutrino physics. In this paper we discussed how the existence of a fourth, mostly sterile neutrino state $\nu_4$ (heavier than the active ones, the so-called $3+1$ model) with a $\nu_e-\nu_s$ mixing characterized by the mixing element $U_{e4}$ would alter the expected $\nu_e$ SN flux at Earth. Obviously, the effect depends on the mass of new state $m_4$ and its mixing $U_{e4}$. However, for a wide range of parameter values (see Eqs.~(\ref{resupp}), (\ref{eq:NHrange}), and (\ref{eq:IHrange})) the  $\nu_e$ radiated from the neutrinosphere convert completely to the $\nu_4$ state en route to the surface of the SN, and hence to the detector at the Earth. Since $\nu_4$ is mostly sterile, this resonant conversion drastically alters the expected early time neutronization $\nu_e$ burst, making it unobservable at leading order. In more detail, due to the small (but not necessarily negligible) mixing $|U_{e4}|^2$, the $\nu_4$ flux has still a chance to be detectable as $\nu_e$ on Earth. However, its kinematic characteristics are altered: depending on the mass $m_4$, energy $E_\nu$ and the distance of SN to Earth, $D$, the $|U_{e4}|^2$-proportional $\nu_e$ flux will be delayed by a time $D(m_4/E_\nu)^2/2c$. We provided an analytical description of the relevant physics, and checked our analytical results (which assume $2\times2$ factorization and adiabaticity of the resonances) against numerical calculations, finding a good agreement. The numerical computations were performed with a $4\times 4$ generalization of the Cayley-Hamilton formalism described in~\cite{Ohlsson:1999xb}, and we report the relevant formulae in Appendix~\ref{4x4CH}.

Our main results can be thus summarized as follows: If the mass hierarchy will be unknown at the time of future Galactic SN detection, the presence of a fourth sterile state can fake the NH phenomenology (lack of observable neutronization burst) even for IH in the active neutrino sector. Turning the argument around, should the active neutrino hierarchy be determined to be of the IH type, the existence of a sterile state may be one of the simplest explanation for  a lack of visible neutronization burst observation from a future SN events in a sufficiently large $\nu_e$ detector. This may corroborate independent evidence from the lab, but also be sensitive to mixing values below current constraints. On the other hand (and perhaps more important), the observation of a neutronization peak consistent with expectations for IH would exclude the existence of a sterile state over a much wider parameter space than what required by laboratory anomalies fits, or even the one testable by detectors coming on-line in the near future. This provides yet another nice example of interplay and complementarity of the astroparticle observables with laboratory ones.

What are the chances that this signature can be actually observed? In the past decade, there have been dedicated studies concerning the detectability of the neutronization burst with different techniques, see notably~\cite{GilBotella:2003sz,Kachelriess:2004ds}. Here we just recall the main results, requirements and challenges, addressing the reader to the original literature for details.  Obviously, the identification of the neutronization burst is especially clean with detectors using the charged-current absorption of $\nu_e$'s. The most widely discussed large detector option for this channel is provided by liquid argon, dominantly via $\nu_e+{}^{40}{\rm Ar}\to e^{-}\,+{}^{40}{\rm K}^*$. The study in~\cite{GilBotella:2003sz}, considering a 70 kton detector, showed that the presence or absence of a neutralization burst leads to a count number of events as different as 86 vs. 41 within the first 240 ms of the signal, for a fiducial SN model located at 10 kpc from us (note that from within 10 kpc  one expects roughly 50\% chances to observe the next SN, see for example the distribution in Ref.~\cite{Mirizzi:2006xx}). Even accounting for Poisson fluctuations and (small) model-to-model variations, in Ref.~\cite{Kachelriess:2004ds} it was estimated that a $2\sigma$ discrimination could be achieved already by this counting test, provided that the distance to the SN is known. Needless to say, a closer SN could allow a separation even with a smaller detector (or equivalently to a higher confidence level, for the benchmark case of 70 kton), while an uncertainty in the position would worsen the sensitivity. Note however that a more refined test exploiting the {\it time structure} might improve the perspectives for diagnostics.

The other technique that has been investigated concerns large water Cherenkov detectors, such as the proposed Hyper-Kamiokande in Japan. This experimental technique is less clean, since the $\nu_e$ elastic scattering on electrons has to compete with other large signals from inverse beta decay on protons, reactions on oxygen and $\nu_x$ scattering onto electrons, but the larger masses (Mton scale) permit to take advantage of the higher statistics. Furthermore, the $\nu_e$ elastic scattering on electrons are more forward peaked and less energetic than most background events. The other channels can be most effectively separated if the detector is doped with Gadolinium, as suggested in~\cite{Beacom:2003nk} and currently tested at the EGADS facility with encouraging results~\cite{EGADS}. Accounting for statistical errors, nuclear cross section uncertainties, supernova model dependence (such as progenitor mass, equation of state) it was found that the capability of distinguishing the two cases are often better than $3\sigma$ for a fiducial SN at 10 kpc, and never worse than $2\sigma$.

In conclusion, testing for the presence of a conventional neutrino burst in the next Galactic SN signal appears within the reach of next generation of underground neutrino detectors, and actually providing a further particle physics motivation to tackle these major experimental enterprises. It is worth noting, however, that it may be possible to obtain a detection of the neutronization peak already with the currently operating IceCube detector at the South Pole. This instrument offers ``only'' a calorimetric light curve via the correlated increase of ``Cherenkov noise" in its detectors (and thus typically via the inverse beta decay reaction), but the statistics is so high that fine time structures can be revealed. It has been shown that in the first 30 ms or so post-bounce ({\it i.e.}, of emergence of a signal on top of the instrumental noise) the two scenarios with/without neutralization peak are markedly different, see e.g. Fig.~11 in~\cite{Abbasi:2011ss}.

The observational perspectives for the other signature (delayed and energy-distorted peak) remain to be studied. Its detection requires at least comparable performances as for the neutronization peak detection, if not superior, plus some luck in the particle physics parameters, such as relatively large mixing angles and large masses. Energy and timing resolution also play a great role. One can envisage in fact to optimize specific strategies to exploit the peculiar time-energy correlation, perhaps extending earlier proposals for the ``active neutrino'' mass measurements from SN signals, see e.g.~\cite{Totani:1998nf}. For sure, the detection of the delayed peak would be a very specific signature of this scenario, but a dedicated analysis is needed to explore the observational perspectives in the allowed parameter space. We note here that other exotic phenomena have been discussed in relation to the prompt neutronization burst: for example, the appearance of the burst in the $\bar{\nu}_e$ channel due to magnetic moments~\cite{Akhmedov:2003fu} or neutrino decay~\cite{Ando:2004qe}. If anything, these possibilities should highlight the importance of large underground detectors for a high-statistics measurement of the neutrino flux(es) from the next Galactic core-collapse SN. 

While being probably the most spectacular one, the alteration of the neutronization burst is not the only manifestation of the presence of sterile neutrinos in the expected neutrino fluxes. We briefly discussed how the antineutrino channel would also be altered, in particular if small mixings of the sterile state with the $\nu_\mu$ or $\nu_\tau$ are present. Most neutrino detectors use the inverse beta decay reaction for detection and thus are sensitive primarily to $\bar{\nu}_e$, so that this channel may offer a more easily accessible diagnostic tool. We showed how large (up to a factor 3!) alterations of the appearance probabilities are induced by the presence of a sterile state. A natural follow-up of our article would be to study the observational signatures of this channel as well, either in the number, energy and time distribution of events in a Water Cherenkov detector, say, or in the luminosity profile that can be measured with impressive detail in a detector like IceCube. Finally, one might wonder if specific signatures of the kinematical time-delay may be inferred from other techniques than the study of the neutronization burst. One possible direction would be to consider if alterations of the time variation of the neutrino emissions revealed in simulations (at ms level, due to anisotropic mass flows in the accretion layer around the newly-formed neutron star) are detectable, along the lines of the study~\cite{Ellis:2012ji} for mass constraints of the active neutrinos.

\begin{acknowledgments}
O.~L.~G.~P. thanks the ICTP and the financial support from the funding grant 2012/16389-1, S\~ao Paulo Research Foundation (FAPESP). A.~E. thanks the financial support from the funding grant 2009/17924-5, S\~ao Paulo Research Foundation (FAPESP) and from the funding grant Jovem Pesquisador from FAEPEX/UNICAMP. P.~S. would like to thank the Instituto de F\'isica Gleb Wataghin at UNICAMP for hospitality during the initial stages of this work and financial support from the funding grant 2012/08208-7, S\~ao Paulo Research Foundation (FAPESP). At LAPTh, this activity was developed coherently with the research axes supported by the Labex grant ENIGMASS. We thank A.~Mirizzi for useful comments on the manuscript.
\end{acknowledgments}

\appendix
\section{Cayley-Hamilton formalism for 4$\times$4 matrix}\label{4x4CH}

In this appendix we provide a few more details on the method for computing the
matrix $\Ss$, in Eq.~(\ref{eq:smatrix}), generalizing the results of~\cite{Ohlsson:1999xb}.

For a constant density medium, after propagation for a distance $L$ in the medium and apart from an overall phase irrelevant for neutrino oscillations, one can write $\Ss(L) = e^{-i\mathcal{H}_m\,L}$, where $\mathcal{H}_m$ is the total Hamiltonian including both the vacuum and the MSW potential terms. By the use of the Cayley--Hamilton formalism, the exponential of $\mathcal{H}_m$ can be rewritten as a simple polynomial in the matrix $T = \mathcal{H}_m - \tr(\mathcal{H}_m){\sf I}/4$, namely the traceless part of the Hamiltonian. In particular, we find
\begin{equation}
\Ss(L)=\sum_{i=1}^4\frac{e^{-i\lambda_i\,L}}{c_1+2c_2\lambda_i+4\lambda_i^3}\left[(c_1+c_2\lambda_i+\lambda_i^3)\,{\sf I}+(c_2+\lambda_i^2)\,{\sf T}+\lambda_i\,{\sf T}^2+\,{\sf T}^3\right]\label{keyeq}
\end{equation}
where $\lambda_i$ are the eigenvalues of $T$, i.e. they are roots of the characteristic equation 
\begin{equation}
\lambda^4+c_2\lambda^2+c_1\lambda+c_0=0\,,
\end{equation}
with the coefficients $c_a$ being
\begin{eqnarray}
&&c_0=\det(T)=\lambda_1\lambda_2\lambda_3\lambda_4\label{c0rel}~,\\
&&c_1=-\tr(T^3)/3=-(\lambda_1\lambda_2\lambda_3+\lambda_1\lambda_2\lambda_4+\lambda_1\lambda_3\lambda_4+\lambda_2\lambda_3\lambda_4)\label{c1rel}~,\\
&&c_2=-\tr(T^2)/2=\lambda_1\lambda_2+\lambda_1\lambda_3+\lambda_1\lambda_4+\lambda_2\lambda_3+\lambda_2\lambda_4+\lambda_3\lambda_4\label{c2rel}\,.
\end{eqnarray}
Note that since $T$ is traceless, the $\lambda_i$'s satisfy 
\begin{equation}
\lambda_1+\lambda_2+\lambda_3+\lambda_4=0\,,\label{c3rel}
\end{equation}
a property which has been used above.

Note that the formulae above apply to a medium of constant density. When neutrinos propagate through a medium of varying density, the electron number density profile can be approximated by a large number $k$ of layers with constant electron number density. If one labels the evolution operator of layer $i$ by $\Ss_i$, then the total evolution operator $\Ss$ is given by
\begin{equation}
\Ss = \Ss_k \Ss_{k-1} \ldots \Ss_2 \Ss_1~.
\label{eq:seq}
\end{equation}

\subsection{Proof of Eq.~(\ref{keyeq})}

First, note that the steps from Eq.~(20) to Eq.~(26) of Ref.~\cite{Ohlsson:1999xb} are generic for a $N\times N$ matrix and thus hold unchanged. Then, the problem is reduced to finding $a_k\,$ ($k=0,\ldots,3$) in the following equation
\begin{equation} 
e^{-i\mathcal{H}_m\,L}=e^{-i\,T\,L}=\sum_{k=0}^3
a_k(-i\,L\,T)^k\,.\label{exponT}
\end{equation} 
In the basis where $T$ is diagonal, the above equation leads to a set of four relations of the type
\begin{equation} 
e^{-i\,\lambda_n\,L}=\sum_{k=0}^3
a_k(-i\,L\,\lambda_n)^k\,,\:\:\:\:n=0,\ldots,3
\end{equation} 
whose inversion leads to the explicit expressions for the $a_k$ (not presented explicitly here). Finally, by plugging the obtained expressions into Eq.~(\ref{exponT}) and grouping the terms proportional to $e^{-i\,\lambda_n\,L}$, one arrives at
\begin{eqnarray}
\Ss(L)&=&\frac{e^{-i\lambda_1\,L}}{(\lambda_2-\lambda_1)(\lambda_3-\lambda_1)(\lambda_4-\lambda_1)}\left[\lambda_2\lambda_3\lambda_4\,{\sf I}-(\lambda_2\lambda_3+\lambda_2\lambda_4+\lambda_3\lambda_4)\,{\sf T}+(\lambda_2+\lambda_3+\lambda_4)\,{\sf T}^2-\,{\sf T}^3\right]+\nonumber\\
&&\frac{e^{-i\lambda_2\,L}}{(\lambda_1-\lambda_2)(\lambda_3-\lambda_2)(\lambda_4-\lambda_2)}\left[\lambda_1\lambda_3\lambda_4\,{\sf I}-(\lambda_1\lambda_3+\lambda_1\lambda_4+\lambda_3\lambda_4)\,{\sf T}+(\lambda_1+\lambda_3+\lambda_4)\,{\sf T}^2-\,{\sf T}^3\right]+\nonumber\\
&&\frac{e^{-i\lambda_3\,L}}{(\lambda_1-\lambda_3)(\lambda_2-\lambda_3)(\lambda_4-\lambda_3)}\left[\lambda_1\lambda_2\lambda_4\,{\sf I}-(\lambda_1\lambda_2+\lambda_1\lambda_4+\lambda_2\lambda_4)\,{\sf T}+(\lambda_1+\lambda_2+\lambda_4)\,{\sf T}^2-\,{\sf T}^3\right]+\nonumber\\
&&\frac{e^{-i\lambda_4\,L}}{(\lambda_1-\lambda_4)(\lambda_2-\lambda_4)(\lambda_3-\lambda_4)}\left[\lambda_1\lambda_2\lambda_3\,{\sf I}-(\lambda_1\lambda_2+\lambda_1\lambda_3+\lambda_2\lambda_3)\,{\sf T}+(\lambda_1+\lambda_2+\lambda_3)\,{\sf T}^2-\,{\sf T}^3\right]\,.
\end{eqnarray}
Note that in the limit of $|\lambda_4|\to \infty$, the above expressions reduce to the $3\times 3$ result explicated in~\cite{Ohlsson:1999xb}.

By using Eq.~(\ref{c3rel}), one has
\begin{eqnarray}
e^{-i\mathcal{H}_mL}&=&\frac{e^{-i\lambda_1\,L}}{(\lambda_2-\lambda_1)(\lambda_3-\lambda_1)(\lambda_4-\lambda_1)}\left[\lambda_2\lambda_3\lambda_4\,{\sf I}-(\lambda_2\lambda_3+\lambda_2\lambda_4+\lambda_3\lambda_4)\,{\sf T}-\lambda_1\,{\sf T}^2-\,{\sf T}^3\right]+\nonumber\\
&&\frac{e^{-i\lambda_2\,L}}{(\lambda_1-\lambda_2)(\lambda_3-\lambda_2)(\lambda_4-\lambda_2)}\left[\lambda_1\lambda_3\lambda_4\,{\sf I}-(\lambda_1\lambda_3+\lambda_1\lambda_4+\lambda_3\lambda_4)\,{\sf T}-\lambda_2\,{\sf T}^2-\,{\sf T}^3\right]+\nonumber\\
&&\frac{e^{-i\lambda_3\,L}}{(\lambda_1-\lambda_3)(\lambda_2-\lambda_3)(\lambda_4-\lambda_3)}\left[\lambda_1\lambda_2\lambda_4\,{\sf I}-(\lambda_1\lambda_2+\lambda_1\lambda_4+\lambda_2\lambda_4)\,{\sf T}-\lambda_3\,{\sf T}^2-\,{\sf T}^3\right]+\nonumber\\
&&\frac{e^{-i\lambda_4\,L}}{(\lambda_1-\lambda_4)(\lambda_2-\lambda_4)(\lambda_3-\lambda_4)}\left[\lambda_1\lambda_2\lambda_3\,{\sf I}-(\lambda_1\lambda_2+\lambda_1\lambda_3+\lambda_2\lambda_3)\,{\sf T}-\lambda_4\,{\sf T}^2-\,{\sf T}^3\right]\,.
\end{eqnarray}
Similarly,  the coefficients of the term proportional to ${\sf T}$ can be isolated by using Eq.~(\ref{c2rel}), then using again Eq.~(\ref{c3rel}) one has
\begin{eqnarray}
e^{-i\mathcal{H}_mL}&=&\frac{e^{-i\lambda_1\,L}}{(\lambda_2-\lambda_1)(\lambda_3-\lambda_1)(\lambda_4-\lambda_1)}\left[\lambda_2\lambda_3\lambda_4\,{\sf I}-(c_2+\lambda_1^2)\,{\sf T}-\lambda_1\,{\sf T}^2-\,{\sf T}^3\right]+\nonumber\\
&&\frac{e^{-i\lambda_2\,L}}{(\lambda_1-\lambda_2)(\lambda_3-\lambda_2)(\lambda_4-\lambda_2)}\left[\lambda_1\lambda_3\lambda_4\,{\sf I}-(c_2+\lambda_2^2)\,{\sf T}-\lambda_2\,{\sf T}^2-\,{\sf T}^3\right]+\nonumber\\
&&\frac{e^{-i\lambda_3\,L}}{(\lambda_1-\lambda_3)(\lambda_2-\lambda_3)(\lambda_4-\lambda_3)}\left[\lambda_1\lambda_2\lambda_4\,{\sf I}-(c_2+\lambda_3^2)\,{\sf T}-\lambda_3\,{\sf T}^2-\,{\sf T}^3\right]+\nonumber\\
&&\frac{e^{-i\lambda_4\,L}}{(\lambda_1-\lambda_4)(\lambda_2-\lambda_4)(\lambda_3-\lambda_4)}\left[\lambda_1\lambda_2\lambda_3\,{\sf I}-(c_2+\lambda_4^2)\,{\sf T}-\lambda_4\,{\sf T}^2-\,{\sf T}^3\right]\,.
\end{eqnarray}
Finally, writing the coefficients of the term proportional to ${\sf I}$ by using Eq.~(\ref{c1rel}) and  then using again iteratively Eq.~(\ref{c2rel}) and Eq.~(\ref{c3rel}), one has
\begin{eqnarray}
e^{-i\mathcal{H}_mL}&=&\frac{e^{-i\lambda_1\,L}}{(\lambda_2-\lambda_1)(\lambda_3-\lambda_1)(\lambda_4-\lambda_1)}\left[-(c_1+c_2\lambda_1+\lambda_1^3)\,{\sf I}-(c_2+\lambda_1^2)\,{\sf T}-\lambda_1\,{\sf T}^2-\,{\sf T}^3\right]+\nonumber\\
&&\frac{e^{-i\lambda_2\,L}}{(\lambda_1-\lambda_2)(\lambda_3-\lambda_2)(\lambda_4-\lambda_2)}\left[-(c_1+c_2\lambda_2+\lambda_2^3)\,{\sf I}-(c_2+\lambda_2^2)\,{\sf T}-\lambda_2\,{\sf T}^2-\,{\sf T}^3\right]+\nonumber\\
&&\frac{e^{-i\lambda_3\,L}}{(\lambda_1-\lambda_3)(\lambda_2-\lambda_3)(\lambda_4-\lambda_3)}\left[-(c_1+c_2\lambda_3+\lambda_3^3)\,{\sf I}-(c_2+\lambda_3^2)\,{\sf T}-\lambda_3\,{\sf T}^2-\,{\sf T}^3\right]+\nonumber\\
&&\frac{e^{-i\lambda_4\,L}}{(\lambda_1-\lambda_4)(\lambda_2-\lambda_4)(\lambda_3-\lambda_4)}\left[-(c_1+c_2\lambda_4+\lambda_4^3)\,{\sf I}-(c_2+\lambda_4^2)\,{\sf T}-\lambda_4\,{\sf T}^2-\,{\sf T}^3\right]\,.
\end{eqnarray}
Applying the same tricks to the denominator, one arrives at Eq.~(\ref{keyeq}).

\section{Derivation of analytical expressions for the flux composition at the Earth\label{FluxEarth}}

In this appendix we derive the coefficients $(c_{ee},c_{xe},c_{se})$ and $(\bar{c}_{ee},\bar{c}_{xe},\bar{c}_{se})$ reported in Tables~\ref{tabI} and \ref{tabII}. The Hamiltonian describing neutrino propagation inside the SN can be written in the flavor basis $(\nu_e,\nu_\mu,\nu_\tau,\nu_s)^T$ as\footnote{we implicitly apply the $U_{23}$ rotation matrix and so $\nu_\mu$ and $\nu_\tau$ states are in the so-called propagation basis. Note that since the fluxes of $\nu_\mu$ and $\nu_\tau$ at production point in SN are equal and also the matter potential difference between $\nu_\mu$ and $\nu_\tau$ is quite small ($V_{\mu\tau}\simeq 10^{-5} V_{CC}$~\cite{Botella:1986wy,Mirizzi:2009td}), the $\mu-\tau$ sector can be rotated arbitrarily.}
\begin{equation}
\mathcal{H} = \frac{UM^2U^\dagger}{2E_\nu}+V = \frac{1}{2E_\nu}\left(\begin{array}{cccc} 
m^2_{ee} &  m^2_{e\mu} &  m^2_{e\tau} & m^2_{es} \\
 m^2_{e\mu} &  m^2_{\mu\mu} &  m^2_{\mu\tau} & m^2_{\mu s} \\
m^2_{e\tau} &  m^2_{\mu\tau} &  m^2_{\tau \tau} & m^2_{\tau s} \\
m^2_{es} &  m^2_{\mu s} &  m^2_{\tau s} & m^2_{ss} \\
\end{array}\right) 
 + V_{CC} \left(\begin{array}{cccc} 
 \frac{1}{2} &  0 &0  &0 \\
 0 & -\frac{1}{2} & 0   & 0 \\
 0& 0  &-\frac{1}{2}   &  0\\
0 & 0  &0    &0 \\
\end{array}\right)~, 
\label{eq:4nu}
\end{equation}
where $m^2_{\alpha\beta}\equiv \left(U M^2 U^{\dagger}\right)_{\alpha\beta}$ are the elements of mass matrix in the flavor basis and $M^2={\rm diag}(0,\Delta m_{21}^2,\Delta m_{31}^2,\Delta m_{41}^2)$. In Eq.~(\ref{eq:4nu}) we assume $Y_e = 1/2$. The same Hamiltonian applies to the case of antineutrino by replacing $V_{CC}\to -V_{CC}$ and $U \to U^\ast$. Deep inside the SN where the matter potential dominates, the Hamiltonian takes the following diagonal form 
\begin{equation}
\mathcal{H} \approx 
 \left(\begin{array}{cccc} 
m^2_{ee}  +\frac{V_{CC}}{2} &  0 &0  &0 \\
 0 & m^2_{\mu\mu} -\frac{V_{CC}}{2} & 0   & 0 \\
 0& 0  &m^2_{\tau \tau} -\frac{V_{CC}}{2}   &  0\\
0 & 0  &0    &m^2_{ss} \\
\end{array}\right)~,
\label{eq:deep-matter}
\end{equation}
and so the flavor eigenstates coincide with the matter eigenstates. However, this correspondence between matter and flavor eigenstates depends on the following alternatives: i) the vanishing or finite value of the active-sterile mixing angles; ii) neutrino or antineutrino channel; iii) normal or inverted hierarchy ordering of active neutrinos (for active-sterile hierarchy we always assume $\Delta m_{41}^2>0$). By knowing this correspondence the flavor oscillation probabilities during the propagation of neutrinos through the SN matter can be calculated in the following way: denoting the initial mass eigenstate fluxes by $F^0_{i}$ ($i=1,2,3,4$), the mass eigenstate fluxes outside the SN, $F_i$, are given by:
\begin{equation}\label{eq:pjumpmatrix}
 \left(\begin{array}{c} 
F_{1} \\
F_{2} \\
F_{3} \\
F_{4} \\
\end{array}\right)= \mathbb{P}(\{p_{\rm jump}\})\cdot
 \left(\begin{array}{c} 
F_{1}^0 \\
F_{2}^0 \\
F_{3}^0 \\
F_{4}^0 \\
\end{array}\right)~,
\end{equation}
where $\mathbb{P}(\{p_{\rm jump}\})$ is an $4\times4$ matrix whose elements depend on the set of jumping probabilities $\{p_{\rm jump}\}$ in various resonance regions along the neutrino propagation in the SN matter. The ``adiabaticity" of neutrino propagation means $\{p_{\rm jump}\}\to 0$; and in this limit we obtain  
\begin{equation}\label{eq:adiabatic}
\lim_{\{p_{\rm jump}\}\to 0} \mathbb{P}(\{p_{\rm jump}\}) = {\sf I} \qquad \Longrightarrow \qquad
 \left(\begin{array}{c} 
F_{1} \\
F_{2} \\
F_{3} \\
F_{4} \\
\end{array}\right)=
 \left(\begin{array}{c} 
F_{1}^0 \\
F_{2}^0 \\
F_{3}^0 \\
F_{4}^0 \\
\end{array}\right)~.
\end{equation}
Due to the long distance between SN and Earth, neutrino mass eigenstates outside the SN propagate decoherently en route to the Earth and so the fluxes of neutrinos in flavor basis at Earth are given by
\begin{equation}
F_{\nu_\alpha} = \sum_{i=1}^{4} |U_{\alpha i}|^2 F_{i}~.  
\end{equation}

The various resonances due to different mass-squared differences occur in neutrino or antineutrino channel depending on the hierarchy of neutrino masses. For each resonance we use the following notation: {\it i}) the resonance due to the $(\Delta m_{21}^2,\theta_{12})$ parameters is called $L$-resonance with the jumping probability $p_L$. The $L$-resonance occurs in the neutrino channel. {\it ii}) the resonance due to $(\Delta m_{31}^2,\theta_{13})$ is called $H$-resonance with the jumping probability $p_H$. This resonance is in the neutrino (antineutrino) channel for NH (IH). {\it iii}) the resonance due to $(\Delta m_{41}^2,\theta_{14})$ is called $H^\prime$-resonance with the jumping probability $p_{H^\prime}$. Since we assume $\Delta m_{41}^2>0$, this resonance occurs in neutrino channel. {\it iv}) the two resonances due to $(\Delta m_{41}^2,\theta_{24})$ and $(\Delta m_{41}^2,\theta_{34})$ occur simultaneously and we call them collectively as $H^{\prime\prime}$-resonance with the jumping probability shown by $p_{H^{\prime\prime}}$. The $H^{\prime\prime}$-resonance occurs in antineutrino channel for $\Delta m_{41}^2>0$. 

\begin{figure}[t!]
\centering
\subfloat[Normal Hierarchy]{
 \includegraphics[trim= 70mm 100mm 70mm 
10mm,clip,width=0.5\textwidth]{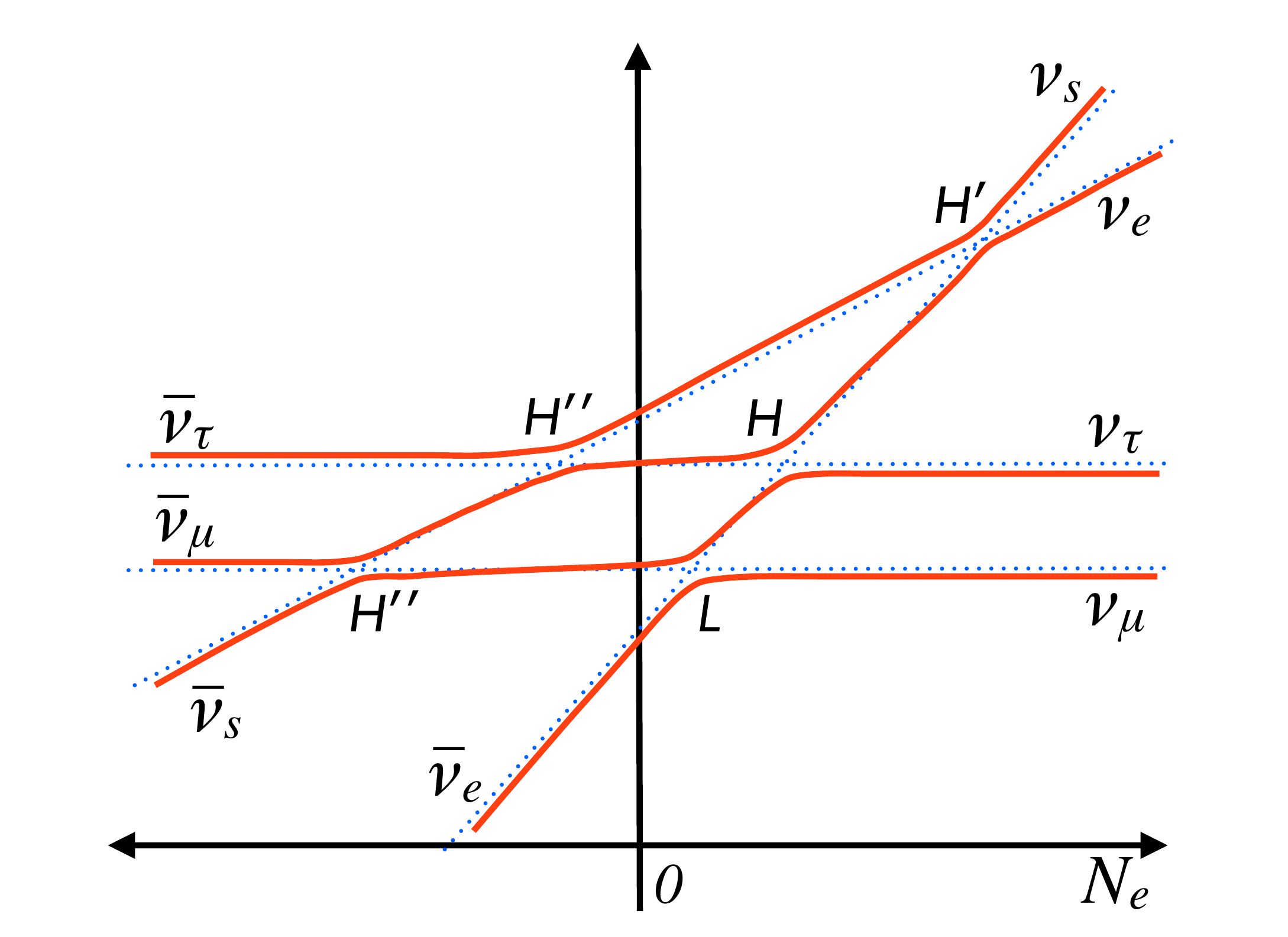}
\label{fig:levelNH}
}
\subfloat[Inverted Hierarchy]{
 \includegraphics[trim= 70mm 103mm 70mm 
10mm,clip,width=0.5\textwidth]{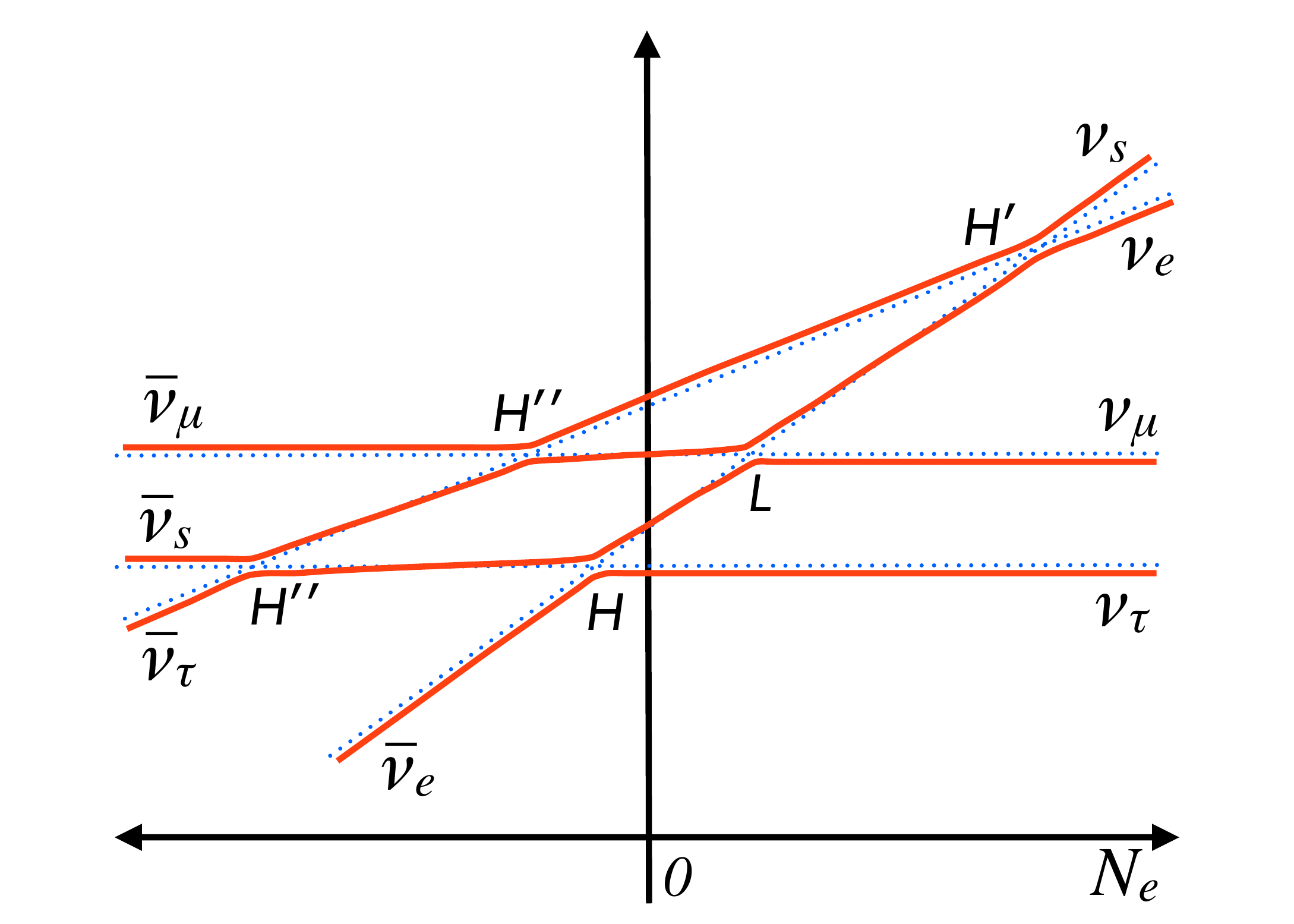}
\label{fig:levelIH}
}
\caption{\label{fig:level}The schematic diagrams of level crossing schemes for NH (left panel) and IH (right panel). In both panels we assume non-vanishing active-sterile mixing angles; {\it i.e.}, $\theta_{i4}\neq0$ for $i=1,2,3$. In the case one (or some) $\theta_{i4}$ vanishes, the corresponding resonance(s) would be ignored.}
\end{figure}

Figure~\ref{fig:level} shows the level crossing diagrams for NH (left panel) and IH (right panel) assuming that all the $\theta_{i4}$ mixing angles are nonzero. The negative values of number density, $N_e$, corresponds to anti-neutrino channel. For the cases where one (or some) of the mixing angles $\theta_{i4}$ vanish, the corresponding resonance region(s) would be ignored. In the following we derive the matrix $\mathbb{P}(\{p_{\rm jump} \})$ and the fluxes $F_{\nu_\alpha}$ for various cases corresponding to vanishing vs. non-vanishing mixing angles, NH vs. IH and neutrino vs. antineutrino channels.

\begin{itemize}
\item{\textsf{Neutrinos, Normal Hierarchy, $\theta_{14}\neq0$}}: In this case, deep inside the SN, the fluxes in flavor basis are related to fluxes in mass basis as (up to a sign)
\begin{equation}\label{eq:case1,fluxes}
 \left(\begin{array}{c} 
F_{1}^0 \\
F_{2}^0 \\
F_{3}^0 \\
F_{4}^0 \\
\end{array}\right)=
 \left(\begin{array}{c} 
F_{\nu_x}^0 \\
F_{\nu_x}^0 \\
F_{\nu_s}^0 \\
F_{\nu_e}^0 \\
\end{array}\right)~.
\end{equation}
The flux of neutrinos in mass basis outside the SN ($F_{i}$) can be obtained by following the level-crossing diagram of the Hamiltonian in Eq.~(\ref{eq:deep-matter}) shown in Figure~\ref{fig:levelNH}. In this case neutrinos pass the $L$, $H$ and $H^{\prime}$-resonances and Eq.~(\ref{eq:pjumpmatrix}) takes the following form
\begin{equation}
 \left(\begin{array}{c} 
F_{1} \\
F_{2} \\
F_{3} \\
F_{4} \\
\end{array}\right)=
 \left(\begin{array}{cccc} 
1 - p_L & (1 - p_H) p_L &  p_{H^\prime} p_H p_L & (1 - p_{H^\prime}) p_H p_L \\
p_L & (1 - p_L) (1 - p_H) & p_{H^\prime} p_H (1 - p_L) & (1 - p_{H^\prime}) p_H (1 - p_L) \\
0 & 0 & 1 - p_{H^\prime} & p_{H^\prime} \\
0 & p_H & p_{H^\prime} (1 - p_H) & (1 - p_{H^\prime}) (1- p_H) \\
\end{array}\right)
 \left(\begin{array}{c} 
F_{1}^0 \\
F_{2}^0 \\
F_{3}^0 \\
F_{4}^0 \\
\end{array}\right)~,
\end{equation}
where obviously in the limit of adiabatic neutrino propagation, {\it i.e.} $(p_L,p_H,p_{H^\prime})\to 0$, the conversion matrix is equal to ${\sf I}$ and Eq.~(\ref{eq:adiabatic}) will be satisfied. Thus, in the adiabatic limit the $\nu_e$ flux at Earth is 
\begin{equation}
F_{\nu_e} = |U_{e4}|^2 F^0_{\nu_e} + \left( |U_{e1}|^2 + |U_{e2}|^2 \right) F^0_{\nu_x} + |U_{e3}|^2 F^0_{\nu_s}~,
\end{equation}
with the $(c_{ee},c_{xe},c_{se})$ coefficients in agreement with Table~\ref{tabI}. This relation is also valid when $\theta_{24}\neq0$ and/or $\theta_{34}\neq0$.

\item{\textsf{Neutrinos, Inverted Hierarchy, $\theta_{14}\neq0$}}: In this case, in the deep SN region we have
\begin{equation}\label{eq:case2,fluxes}
 \left(\begin{array}{c} 
F_{1}^0 \\
F_{2}^0 \\
F_{3}^0 \\
F_{4}^0 \\
\end{array}\right)=
 \left(\begin{array}{c} 
F_{\nu_x}^0 \\
F_{\nu_s}^0 \\
F_{\nu_x}^0 \\
F_{\nu_e}^0 \\
\end{array}\right)~.
\end{equation}
Since the hierarchy is inverted, the $H$-resonance is in the antineutrino channel and neutrinos pass the $L$ and $H^{\prime}$-resonances (see Figure~\ref{fig:levelIH}). In this case Eq.~(\ref{eq:adiabatic}), and the matrix $\mathbb{P}$ in it, can be written in the following way
\begin{equation}
 \left(\begin{array}{c} 
F_{1} \\
F_{2} \\
F_{3} \\
F_{4} \\
\end{array}\right)=
 \left(\begin{array}{cccc} 
1 - p_L & p_{H^\prime} p_L &  0 & p_L \\
0 & 1 - p_{H^\prime} & 0 & p_{H^\prime} \\
0 & 0 & 1 & 0 \\
p_L & p_{H^\prime} (1 - p_L) & 0 & (1 - p_{H^\prime}) (1- p_L) \\
\end{array}\right)
 \left(\begin{array}{c} 
F_{1}^0 \\
F_{2}^0 \\
F_{3}^0 \\
F_{4}^0 \\
\end{array}\right)~.
\end{equation}
Again in the adiabatic limit $\mathbb{P}\to{\sf I}$ and we obtain 
\begin{equation}
F_{\nu_e} = |U_{e4}|^2 F^0_{\nu_e} + \left( |U_{e1}|^2 + |U_{e3}|^2 \right) F^0_{\nu_x} + |U_{e2}|^2 F^0_{\nu_s}~,
\end{equation}
in agreement with Table~\ref{tabI}. This relation is valid also for $\theta_{24}\neq0$ and/or $\theta_{34}\neq0$. 

\item{\textsf{Antineutrinos, Normal Hierarchy, $\theta_{14}\neq0$ and $\theta_{24}=\theta_{34}=0$}}:
In this case the mass and flavor fluxes in the deep SN medium are related by 
\begin{equation}\label{eq:case3,fluxes}
 \left(\begin{array}{c} 
F_{\overline{1}}^0 \\
F_{\overline{2}}^0 \\
F_{\overline{3}}^0 \\
F_{\overline{4}}^0 \\
\end{array}\right)=
 \left(\begin{array}{c} 
F_{\overline{\nu}_e}^0 \\
F_{\overline{\nu}_x}^0 \\
F_{\overline{\nu}_x}^0 \\
F_{\overline{\nu}_s}^0 \\
\end{array}\right)~.
\end{equation}
Since $\theta_{24}=\theta_{34}=0$ and hierarchy is normal there is no resonance in the antineutrino channel and so $\mathbb{P}={\sf I}$. For $\overline{\nu}_e$ flux at the Earth we obtain
\begin{equation}
F_{\overline{\nu}_e} = |U_{e1}|^2 F^0_{\overline{\nu}_e} + \left( |U_{e2}|^2 + |U_{e3}|^2 \right) F^0_{\overline{\nu}_x} + |U_{e4}|^2 F^0_{\overline{\nu}_s}~,
\end{equation}
in agreement with Table~\ref{tabII}.

\item{\textsf{Antineutrinos, Inverted Hierarchy, $\theta_{14}\neq0$ and $\theta_{24}=\theta_{34}=0$}}:
Mass and flavor basis fluxes are related by
\begin{equation}\label{eq:case4,fluxes}
 \left(\begin{array}{c} 
F_{\overline{1}}^0 \\
F_{\overline{2}}^0 \\
F_{\overline{3}}^0 \\
F_{\overline{4}}^0 \\
\end{array}\right)=
 \left(\begin{array}{c} 
F_{\overline{\nu}_x}^0 \\
F_{\overline{\nu}_x}^0 \\
F_{\overline{\nu}_e}^0 \\
F_{\overline{\nu}_s}^0 \\
\end{array}\right)~.
\end{equation}
In this case the $H$-resonance is in the antineutrino channel and the $\mathbb{P}$ matrix is similar to the one for $3\nu$ framework and IH. The Eq.~(\ref{eq:pjumpmatrix}) writes in this case as
\begin{equation}
 \left(\begin{array}{c} 
F_{\overline{1}} \\
F_{\overline{2}} \\
F_{\overline{3}} \\
F_{\overline{4}} \\
\end{array}\right)=
 \left(\begin{array}{cccc} 
1 & 0 &  0 & 0 \\
0 & 1 - p_{H} & p_H & 0 \\
0 & p_H & 1 - p_H & 0 \\
0 & 0 & 0 & 1 \\
\end{array}\right)
 \left(\begin{array}{c} 
F_{\overline{1}}^0 \\
F_{\overline{2}}^0 \\
F_{\overline{3}}^0 \\
F_{\overline{4}}^0 \\
\end{array}\right)~.
\end{equation}
The $\overline{\nu}_e$ flux at the Earth is given by
\begin{equation}
F_{\overline{\nu}_e} = |U_{e3}|^2 F^0_{\overline{\nu}_e} + \left( |U_{e1}|^2 + |U_{e2}|^2 \right) F^0_{\overline{\nu}_x} + |U_{e4}|^2 F^0_{\overline{\nu}_s}~,
\end{equation}
in agreement with Table~\ref{tabII}.

\item{\textsf{Antineutrinos, Normal Hierarchy, $\theta_{24}\neq0$ and/or $\theta_{34}\neq0$}}:
In this case we have
\begin{equation}\label{eq:case5,fluxes}
 \left(\begin{array}{c} 
F_{\overline{1}}^0 \\
F_{\overline{2}}^0 \\
F_{\overline{3}}^0 \\
F_{\overline{4}}^0 \\
\end{array}\right)=
 \left(\begin{array}{c} 
F_{\overline{\nu}_e}^0 \\
F_{\overline{\nu}_s}^0 \\
F_{\overline{\nu}_x}^0 \\
F_{\overline{\nu}_x}^0 \\
\end{array}\right)~.
\end{equation}
The only resonance in the antineutrino channel is the $H^{\prime\prime}$-resonance (see Figure~\ref{fig:levelNH}). The $H^{\prime\prime}$-resonance can originate from nonzero $\theta_{24}$ and/or $\theta_{34}$. For example, for $\theta_{24}\neq0$, Eq.~(\ref{eq:pjumpmatrix}) takes the following form
\begin{equation}\label{eq:p1}
 \left(\begin{array}{c} 
F_{\overline{1}} \\
F_{\overline{2}} \\
F_{\overline{3}} \\
F_{\overline{4}} \\
\end{array}\right)=
 \left(\begin{array}{cccc} 
1 & 0 &  0 & 0 \\
0 & 1 - p_{H^{\prime\prime}} & p_{H^{\prime\prime}} & 0 \\
0 & p_{H^{\prime\prime}} & 1 - p_{H^{\prime\prime}} & 0 \\
0 & 0 & 0 & 1 \\
\end{array}\right)
 \left(\begin{array}{c} 
F_{\overline{1}}^0 \\
F_{\overline{2}}^0 \\
F_{\overline{3}}^0 \\
F_{\overline{4}}^0 \\
\end{array}\right)~,
\end{equation}
and again in the adiabatic limit ($\mathbb{P}={\sf I}$) we obtain
\begin{equation}\label{eq:flux5}
F_{\overline{\nu}_e} = |U_{e1}|^2 F^0_{\overline{\nu}_e} + \left( |U_{e3}|^2 + |U_{e4}|^2 \right) F^0_{\overline{\nu}_x} + |U_{e2}|^2 F^0_{\overline{\nu}_s}~,
\end{equation}
in agreement with Table~\ref{tabII}. The $\mathbb{P}$ matrix in Eq.~(\ref{eq:p1}) changes when the $H^{\prime\prime}$-resonance originates from $\theta_{34}\neq0$; but, however, in the adiabatic limit again $\mathbb{P}={\sf I}$ and the same $\overline{\nu}_e$ flux as in Eq.~(\ref{eq:flux5}) is applicable. Also, this relation holds for both $\theta_{14}=0$ and $\theta_{14}\neq0$.

\item{\textsf{Antineutrinos, Inverted Hierarchy, $\theta_{24}\neq0$ and/or $\theta_{34}\neq0$}}:
Finally, in this case we have
\begin{equation}\label{eq:case6,fluxes}
 \left(\begin{array}{c} 
F_{\overline{1}}^0 \\
F_{\overline{2}}^0 \\
F_{\overline{3}}^0 \\
F_{\overline{4}}^0 \\
\end{array}\right)=
 \left(\begin{array}{c} 
F_{\overline{\nu}_s}^0 \\
F_{\overline{\nu}_x}^0 \\
F_{\overline{\nu}_e}^0 \\
F_{\overline{\nu}_x}^0 \\
\end{array}\right)~.
\end{equation}
Since the hierarchy is inverted antineutrinos pass the $H$ and $H^{\prime\prime}$-resonances (see Figure~\ref{fig:levelIH}). Assuming $\theta_{24}\neq0$ and $\theta_{34}=0$, the $\mathbb{P}$ matrix takes the form
\begin{equation}\label{eq:p2}
 \left(\begin{array}{c} 
F_{\overline{1}} \\
F_{\overline{2}} \\
F_{\overline{3}} \\
F_{\overline{4}} \\
\end{array}\right)=
 \left(\begin{array}{cccc} 
(1 - p_{H^{\prime\prime}}) (1 - p_H) & p_{H^{\prime\prime}} (1 - p_H) &  p_H & 0 \\
p_{H^{\prime\prime}} & 1 - p_{H^{\prime\prime}} & 0 & 0 \\
(1 - p_{H^{\prime\prime}}) p_H & p_{H^{\prime\prime}} p_H & 1 - p_{H} & 0 \\
0 & 0 & 0 & 1 \\
\end{array}\right)
 \left(\begin{array}{c} 
F_{\overline{1}}^0 \\
F_{\overline{2}}^0 \\
F_{\overline{3}}^0 \\
F_{\overline{4}}^0 \\
\end{array}\right)~,
\end{equation}
and so (in the adiabatic limit)
\begin{equation}
F_{\overline{\nu}_e} = |U_{e3}|^2 F^0_{\overline{\nu}_e} + \left( |U_{e2}|^2 + |U_{e4}|^2 \right) F^0_{\overline{\nu}_x} + |U_{e1}|^2 F^0_{\overline{\nu}_s}~,
\end{equation}
in agreement with Table~\ref{tabII}. This relation holds for both $\theta_{14}=0$ and $\theta_{14}\neq0$.

\end{itemize}


\end{document}